\documentclass[iop,revtex4,12pt,preprint]{aa}

\usepackage{graphics,graphicx}
\usepackage{helvet}
\usepackage{subfigure}
\usepackage[utf8]{inputenc}
\usepackage[T1]{fontenc}
\usepackage{soul} 
\usepackage[colorlinks,urlcolor=blue,citecolor=blue,linkcolor=blue,pdfusetitle]{hyperref}
\usepackage{amsmath, amssymb, gensymb}
\usepackage{morefloats}
\usepackage{float}
\usepackage{ulem}
\usepackage{url}
\usepackage{natbib}
\usepackage{microtype}
\usepackage{multirow}
\usepackage{morefloats}
\usepackage{rotating}
\usepackage{here}
\usepackage{xspace}
\usepackage{color}
\usepackage{lipsum}
\usepackage[a4paper]{geometry}
\usepackage{xspace}
\usepackage{tabularx}
\usepackage{supertabular}
\usepackage{caption}
\usepackage{float}
\usepackage{stmaryrd}

\def\eg {e.g.,\xspace} 
\def\ie {i.e.,\xspace} 

\def\kms {km\,s$^{-1}$\xspace}

\def\co {CO\,($J$\,=\,2\,$\rightarrow\,$1)\xspace}

\begin{document}

\title{Physical conditions for dust grain alignment in Class 0 protostellar cores}
\subtitle{I. Observations of dust polarization and molecular irradiation tracers}

\author{V. J. M. Le Gouellec\inst{1,}\inst{2,}\inst{3} , A. J. Maury\inst{1,}\inst{4}, C. L. H. Hull \inst{5,}\inst{6,}\inst{7}}

\institute{
Laboratoire AIM, Paris-Saclay, CEA/IRFU/SAp - CNRS - Université Paris Diderot, 91191 Gif-sur-Yvette Cedex, France
\and
SOFIA Science Center, Universities Space Research Association, NASA Ames Research Center, Moffett Field, California 94035, USA
\email{valentin.j.legouellec@nasa.gov}
\and
European Southern Observatory, Alonso de C\'ordova 3107, Vitacura, Casilla 19001, Santiago , Chile
\and
Harvard-Smithsonian Center for Astrophysics, Cambridge, MA 02138, USA
\and
National Astronomical Observatory of Japan, NAOJ Chile, Alonso de Córdova 3788, Office 61B, 7630422, Vitacura, Santiago, Chile
\and
Joint ALMA Observatory, Alonso de Córdova 3107, Vitacura, Santiago, Chile
\and
NAOJ Fellow
}

\date{}

\abstract
{High angular resolution observations of Class 0 protostars have produced detailed maps of the polarized dust emission in the envelopes of these young embedded objects. Interestingly, the improved sensitivity brought by ALMA has   revealed wide dynamic ranges of polarization fractions, with specific locations harboring surprisingly large amounts of polarized dust emission.}
{Our  aim is to characterize the grain alignment conditions and dust properties responsible for the observed polarized dust emission in the inner envelopes ($\leq\,1000$ au) of Class 0 protostars.}
{We analyzed the polarized dust emission maps obtained with ALMA and compared them to molecular line emission maps of specific molecular tracers, mainly C$_2$H, which allowed us to probe one of the key components in dust grain alignment theories: the irradiation field.}
{We show that C$_2$H peaks toward outflow cavity walls, where the polarized dust emission is also enhanced. Our analysis provides a tentative correlation between the morphology of the polarized intensity and C$_2$H emission, suggesting that the radiation field impinging on the cavity walls favors both the grain alignment and the warm carbon chain chemistry in these regions. 
We propose that shocks happening along outflow cavity walls could potentially represent an additional source of photons contributing to dust grain alignment.
However, some parts of the cores, such as the equatorial planes, exhibit enhanced polarized flux,  although no radiation driven chemistry is observed, for example where radiative torques are theoretically not efficient enough. This suggests that additional physical conditions, such as source geometry and dust grain evolution, may play a role in grain alignment.
}
{Comparing chemical processes with grain alignment physics opens a promising avenue to develop our understanding of the dust grain evolution (\ie their origin, growth, and structure) in the interior of Class 0 protostars. The source geometry and evolution can represent important factors that set the environmental conditions of the inner envelope, determining whether the radiation field strength and spectrum can drive efficient dust grain alignment via radiative torques.
}


\titlerunning{Constraining the radiation field via dust polarization}
\authorrunning{Le Gouellec et al.}

\maketitle

\section{Introduction}

Located in the densest parts of molecular clouds, Class 0 protostars are the youngest low-mass star forming objects \citep{Andre1993,Andre1994}. They are composed of a growing protostellar embryo at the center, which accretes material from the surrounding cold protostellar envelope.  These objects have yet to accrete the main part of their future main sequence star mass as most of their mass is still located in the envelope, such that $L_{\textrm{sub-mm}}/L_{\textrm{bol}}\gg0.5\%$ and $M_{\textrm{env}}\gg M_{\star}$. The Class 0 protostellar phase thus corresponds to a period of vigorous accretion activity whose physical characteristics will set the properties of the future star and protoplanetary disk. At this stage the accretion is ruled by a variety of physical processes, and it remains unclear if, for example, the accretion is disk-regulated, as it is in later stage T Tauri stars \citep{Hartmann2016}, or if it occurs directly from the infalling flows of the envelope \citep{LeeYN2021}. The accretion is responsible for the vigorous ejection of material in the form of (bi)polar jets \citep{Bachiller1999,Arce2007,Bally2016}, and for an important irradiation feedback on the envelope caused by the accretion shock onto the central protostellar embryo \citep{Spaans1995}. Outflow cavities surrounding the jets are also formed, potentially powered by the collimated jet \citep{Raga1993} and/or by the ejection of low-velocity winds launched at disk scales \citep{Louvet2018,deValon2020}.

Outflow cavities are directly subject to the radiation field emanating from the accretion activity \citep{Krumholz2012}, which may highly influence the chemistry of the inner envelope \citep{Stauber2004,Stauber2005}. For instance, far-infrared rotational lines of CO and H$_2$O from the bulk of the emission of young protostars \citep{Spaans1995,Hogerheijde1998,vanKempen2009,Kristensen2013b}, were reproduced by models accounting for UV-heated gas and/or shocks along outflow cavity walls \citep{Goicoechea2012,Visser2012,LeeS2015}. The UV photons coming from the accretion disk around the central protostellar embryo, which are reprocessed or are scattered into cavities, are responsible for heating the gas and thus creating photon dominated region (PDR) conditions in outflow cavity walls, which can   also be probed by [CI] and [CII] lines; the presence of these lines suggests the presence of UV-irradiated shocks \citep{vanKempen2009b,Goicoechea2012,Yildiz2012}. Farther  away from the central object, additional UV photons can be produced in bow shocks of the outflow \citep{Boehm1993}, if $J$-shocks are present \citep{Neufeld1989a,Neufeld1989b}, and/or by small-scale shocks along the walls of the cavities \citep{Curiel1995,Saucedo2003,Walter2003}. The presence of irradiating photons was also confirmed by observations along the outflow cavity walls of molecules tracing warm chemistry processes involving UV photons, such as C$_2$H and $c$-C$_3$H$_2$ \citep{Murillo2018,Tychoniec2021}, because the formation pathway of these molecules requires a significant amount of UV irradiation. The corresponding chemistry is similar to what is encountered at the edge of PDR zones, for example the Horsehead PDR \citep{Cuadrado2015,Guzman2014,Guzman2015}.

On the other hand, recent observations have revealed that outflow cavity walls also  seem   to harbor specific grain alignment conditions as strong polarized dust emission along the walls of the bipolar outflow cavities were detected \citep{Hull2017b,Cox2018,Maury2018,Ko2019,Kwon2019,LeGouellec2019a,Hull2020a}. In regions of optically thin thermal dust emission, which is the case for the typical spatial scales resolved by ALMA toward protostellar envelopes (\ie $\sim$ 50 to 1000 au), the polarization is caused by the thermal emission of aligned and aspherical dust grains. In the typical conditions prevailing in protostellar envelopes, the most likely grain alignment process is caused by   radiative alignment torques (RATs; \citealt{Draine1996,Draine1997,LazarianHoang2007,Andersson2015}), which aligns the dust grains' minor axis parallel to the ambient magnetic field lines. This causes  the linear polarization of the dust thermal emission to be orthogonal to the magnetic field component projected on the plane of the sky. The RAT mechanism is subject to the local environmental conditions (\ie the mean wavelength, strength, and degree of anisotropy of the radiation field), the dust characteristics (size, shape, composition), and the gas temperature and density. Understanding the causes for the strong polarized dust emission detected in outflow cavity walls and within the inner envelope ($\leq 500 $ au) is thus crucial to further developing the knowledge we have of the impacts that accretion has on what surrounds the central protostellar embryo. 
Because it governs the gaseous collisional rate that randomizes the orientation of the dust grains,   gas density is one of the key parameters that set the grain alignment conditions (see, \eg \  \citealt{Reissl2020}). However, the polarization fraction maps are irregular with  nonhomogeneous spatial distribution in protostellar envelopes. The high values exhibited in outflow cavity walls may point toward the total intensity not being a good proxy for the local gas volume  density and/or specific grain alignment conditions such as favored irradiation.
In this work, divided into two papers, we make a joint effort to analyze both the UV-sensitive chemistry and the efficiency of the dust grain alignment in the interiors of Class 0 protostars. 
This paper focuses on comparing the location of strongly irradiated molecular gas with the spatial distribution of aligned grains in the envelope of Class 0 protostellar cores through tracers of warm chemistry and polarized dust emission. 
In this first work, which relates to observations alone, we use preferentially the polarized intensity as a proxy for the morphology of the favorable grain alignment conditions to qualitatively address the role of the radiation field.
The polarization fraction values taken into account with the levels of magnetic field organization will be discussed in Paper II, which will focus on reproducing the observed grain alignment efficiency via comparisons with radiative transfer modeling, where the role of the radiation field is investigated.

This paper is structured as follows. In Section \ref{sec:p3_obs} we present ALMA dust polarization observations alongside ALMA molecular line observations of the UV-sensitive C$_2$H molecule and the shock-tracer SO  in a sample of Class 0 protostars.   In Section \ref{sec:p3_obs_KS} the spatial distribution of polarized dust emission is qualitatively compared with the distributions of the molecular emission lines. Finally, in Section \ref{sec:p3_disc} we discuss the comparisons between the dust polarization and irradiation tracer observational results. We draw our conclusions in Section \ref{sec:p3_ccl}.

\section{ALMA observations of the polarized dust emission and molecular gas tracers}
\label{sec:p3_obs}

\begin{table*}[!tbph]
\centering
\small
\caption[ALMA dust polarization and molecular line observations details]{Sources and ALMA dust polarization observations details}
\label{t.obs_pol}
\setlength{\tabcolsep}{0.2em} 
\begin{tabular}{p{0.21\linewidth}cccccccc}
\hline \hline \noalign{\smallskip}
Name & $\alpha_{\textrm{J2000}}$ & $\delta_{\textrm{J2000}}$ & $M_\textrm{env}$$^a$ & $L_{\textrm{bol}}$$^a$&  Dist.$^b$& $\lambda$$^c$ & $\theta_{\textrm{res}}$$^d$ & Reference$^e$ \\
&&&$M_\odot$&$L_\odot$&pc&(mm)& (au)&\\
\noalign{\smallskip}  \hline
\noalign{\smallskip}
B335 & 19:37:00.91 & \phantom{-}07:34:09.60 & \phantom{0}1.44 & \phantom{00}2.7 & 165 & 1.3\phantom{0} & 101 & \citet{Maury2018}, in prep. \\
\noalign{\smallskip}
\hline
\noalign{\smallskip}
L1448 IRS2 & 03:25:22.41 & \phantom{-}30:45:13.21 & \phantom{0}1.3\phantom{0} & \phantom{00}7.0 & 294 & 1.3\phantom{0} & 135 & \citet{Kwon2019} \\
\noalign{\smallskip}
\hline
\noalign{\smallskip}
NGC1333 IRAS4A &03:29:10.55&\phantom{-}31:13:31.00 &12.3\phantom{0}&\phantom{0}14.2 & 294 & 1.3\phantom{0} & 112 & \citet{Ko2019} \\
\noalign{\smallskip}
\hline
\noalign{\smallskip}
Serpens Emb 8  &18:29:48.09 & \phantom{-}01:16:43.30 & \multirow{2}{*}{12.8\phantom{0}}&\multirow{2}{*}{\phantom{00}7.3} & 484 & 0.87 & 161 & \citet{Hull2017a}\\
Serpens Emb 8(N) & 18:29:48.73 & \phantom{-}01:16:55.61 &&& 484 & 0.87 & 161 &  \citet{LeGouellec2019a}  \\
\noalign{\smallskip}
\hline
\smallskip
\end{tabular}
\vspace{-0.15in}
\caption*{\small
For the CCH spectral lines presented in the figures below, we integrated over the two hyperfine structures in order to construct the moment 0 maps.\\
$^a$ Envelope mass $M_\textrm{env}$ and bolometric luminosity $L_{\textrm{bol}}$ values. The values calculated for Serpens Emb 8 and Emb 8(N) include the two sources together \citep{Enoch2009b,Enoch2011}. \citet{Kurono2013,Maury2018} for B335. \citet{Sadavoy2014,Karska2018} for L1448 IRS2. \citet{Karska2018,Galametz2019} for IRAS4A. \\
$^b$ Distance reported in the literature. \citet{Zucker2019} for  Perseus and Serpens. \citet{Watson2020} for B335. We scale the masses and luminosities found in the literature to the distances listed here.\\
$^c$ Wavelength of dust polarization observations.\\
$^d$ Angular resolution in au. We took the effective synthesized beam size of the ALMA maps.\\
$^e$ Reference of the publication(s) presenting the ALMA dust polarization dataset(s).
}
\end{table*}

\begin{table*}[!tbph]
\centering
\small
\caption[ALMA molecular line observations details]{ALMA molecular line observations details}
\label{t.obs_lines}
\setlength{\tabcolsep}{0.2em} 
\begin{tabular}{p{0.21\linewidth}cccccc}
\hline \hline \noalign{\smallskip}
Name  &Spectral Lines $^a$&$\nu$$^b$ & Reference$^c$\\
&&(GHz)&\\
\noalign{\smallskip}  \hline
\noalign{\smallskip}
 \multirow{4}{*}{B335} & C$_2$H\;($J_{N,F}$=7/2$_{3,4}$–5/2$_{2,3}$) &262.004260 &  \multirow{4}{*}{\citet{Imai2016}}\\
& C$_2$H\;($J_{N,F}$=7/2$_{3,3}$–5/2$_{2,2}$) &262.006482&\\
&CS\;($J$=5–4)&244.935644&\\
& $c$-C$_3$H$_2$\;(5$_{2,3}$–4$_{3,2}$)&249.054368&\\
\noalign{\smallskip}
\hline
\noalign{\smallskip}
 \multirow{3}{*}{L1448 IRS2 }&  C$_2$H\;($J_{N,F}$=5/2$_{3,3}$–3/2$_{2,2}$) &262.064986 &  \multirow{3}{*}{Zhang et al. in prep.}\\
& C$_2$H\;($J_{N,F}$=5/2$_{3,2}$–3/2$_{2,1}$) &262.067469&\\
&CS\;($J$=5–4)&244.935644&\\
\noalign{\smallskip}
\hline
\noalign{\smallskip}
 \multirow{4}{*}{NGC1333 IRAS4A} & C$_2$H\;($J_{N,F}$=7/2$_{3,4}$–5/2$_{2,3}$) &262.004260& \multirow{4}{*}{\citet{Chuang2021}}\\
& C$_2$H\;($J_{N,F}$=7/2$_{3,3}$–5/2$_{2,2}$) &262.006482&\\
&CS\;($J$=5-4)&244.935644&\\
&SO\;($N_J$=7(6)–6(5))&261.843684&\\
\noalign{\smallskip}
\hline
\noalign{\smallskip}
\multirow{3}{*}{Serpens Emb 8 } & C$_2$H\;($J_{N,F}$=7/2$_{3,4}$–5/2$_{2,3}$) &262.004260  &\multirow{3}{*}{\citet{vanGelder2020}}\\
& C$_2$H\;($J_{N,F}$=7/2$_{3,3}$–5/2$_{2,2}$) &262.006482&\\
&SO\;($N_J$=7(6)–6(5))&261.843684&\\
\noalign{\smallskip}
\multirow{3}{*}{Serpens Emb 8(N)} & C$_2$H\;($J_{N,F}$=7/2$_{3,4}$–5/2$_{2,3}$) &262.004260 & \multirow{2}{*}{\citet{LeGouellec2019a}} \\
& C$_2$H\;($J_{N,F}$=7/2$_{3,3}$–5/2$_{2,2}$) &262.006482&\\
&SO\;($N_J$=6(5)–5(4))&219.949433& Tychoniec et al., in prep.\\
\noalign{\smallskip}
\hline
\smallskip
\end{tabular}
\vspace{-0.15in}
\caption*{\small
For the CCH spectral lines presented in the figures below, we integrated over the two hyperfine structures in order to construct the moment 0 maps.\\
$^a$ Spectral lines that we present in this study.\\
$^b$ Frequency of spectral line observations. \\
$^b$ Reference of the publication(s) presenting the ALMA molecular line dataset(s).
}
\end{table*}

\begin{figure*}[!tbh]
\centering
\hspace{-0.7cm}
\subfigure{
\includegraphics[scale=0.6227,clip,trim= -0.5cm 0cm 0.3cm 0.15cm]{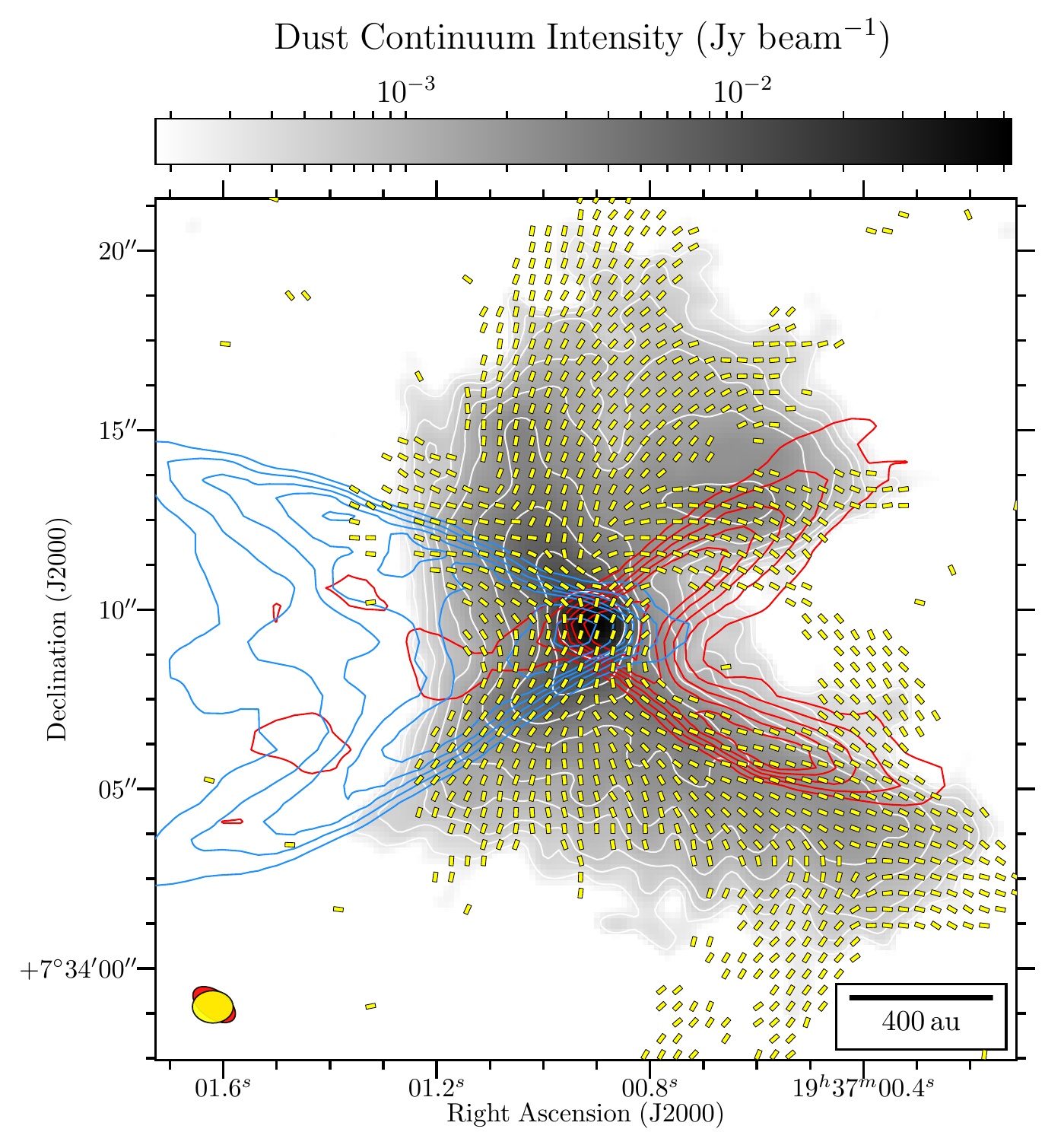}}
\subfigure{
\includegraphics[scale=0.616,clip,trim= 2cm 2cm 4cm 2cm]{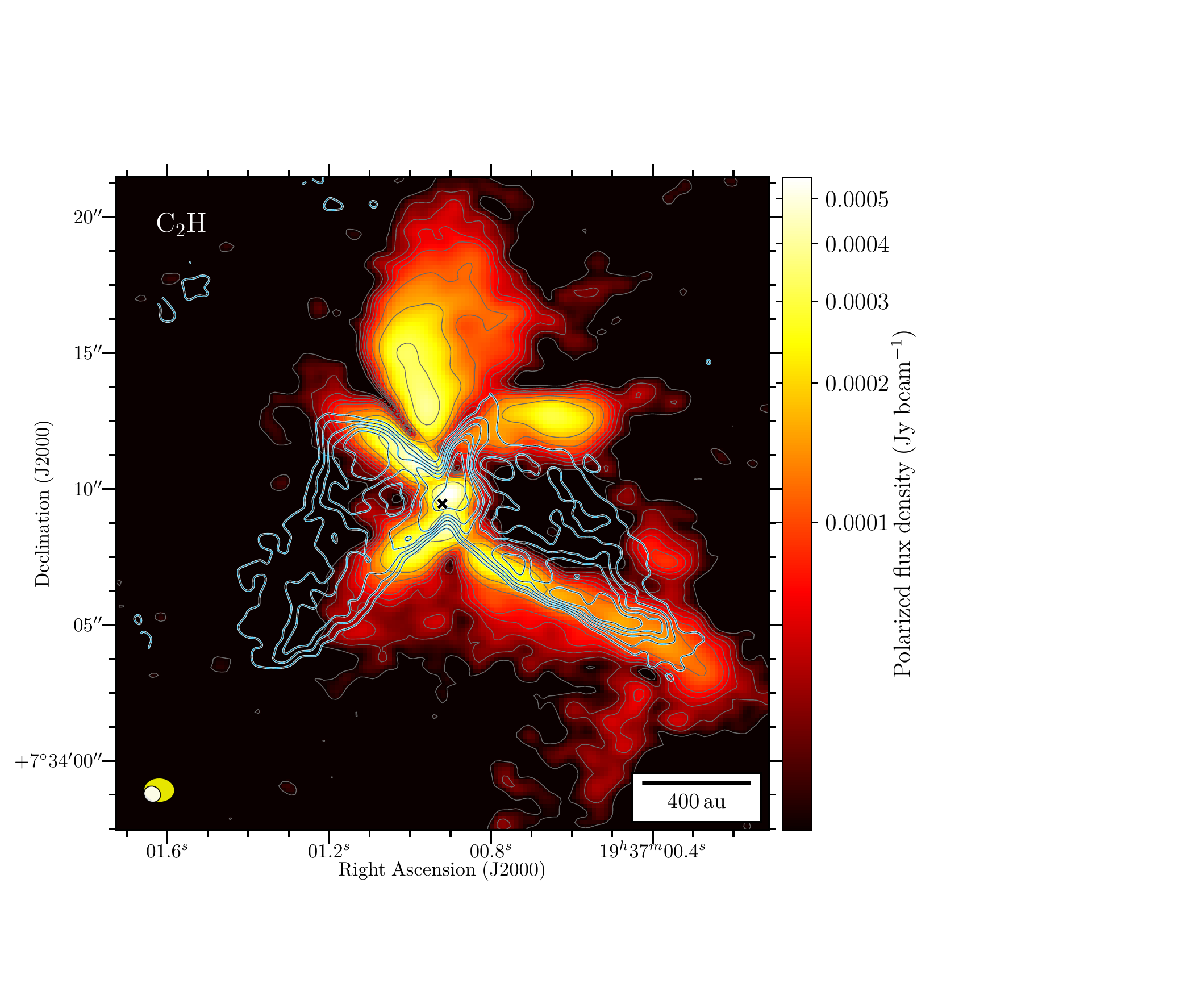}}
\vspace{-0.4cm}
\caption[Magnetic fields, polarized intensity, and molecular line observations around the B335 Class 0 protostellar core]{\footnotesize Observations of  magnetic fields, polarized intensity, and molecular lines  around the B335 Class 0 protostellar core. \textit{Left panel:} Line segments represent magnetic field lines, rotated by 90$^{\circ}$ from the dust polarization angle (the length of a segment does not represent a quantity). They are plotted where the polarized intensity $P\,>\,3\sigma_{P}$. The gray  scale is the total intensity (Stokes $I$) of the thermal dust emission, shown from 3$\sigma_I$. The white contours trace the total intensity at levels of 5, 8, 11, 16, 24, 44, 74, 128, 256 $\times\,\sigma_I$. The blue and red contours trace the moment 0 map of the blueshifted (by integrating emission from –25 to 7 \kms) and redshifted (by integrating emission from 10 to 35 \kms) \co map at levels of 3, 5, 7, 9, 11, 16, 24, 44, 74, 128, 256, 400 $\times$ the rms noise level of the map. \textit{Right panel:} In each panel the color scale is the dust polarization intensity, shown from 3$\sigma_P$. The crosses indicate the peak of the dust continuum map. The thin gray contours trace the dust polarization intensity, at levels of 3, 5, 7, 9, 11, 16, 24, 44 $\times\,\sigma_P$. The thick light blue contours trace the moment 0 map of the C$_2$H (by integrating emission from 7 to 12.5 \kms) molecular emission spectral line (see Table \ref{t.obs_lines} for details of the line transitions) at levels of 3, 5, 7, 9, 11, 16, 24, 44 $\times$ the rms noise level of each map. In the  bottom left corner of each panel the ellipses represent the beam resolution element of a given dataset: yellow for polarized dust emission, red for CO bipolar outflow emission, and white for the CCH spectral emission line. The ALMA 1.3mm dust polarized emission is from \citet{Maury2018} and Maury et al. (in preparation). The ALMA \co bipolar outflow emission is from \citet{Cabedo2021}. The C$_2$H emission line is from \citet{Imai2016}.
}
 \label{fig:obs_b335_chem}
\end{figure*}

\begin{figure*}[!tbh]
\centering
\hspace{-0.7cm}
\subfigure{
\includegraphics[scale=0.625,clip,trim=-0.7cm 0cm 0.5cm 0cm]{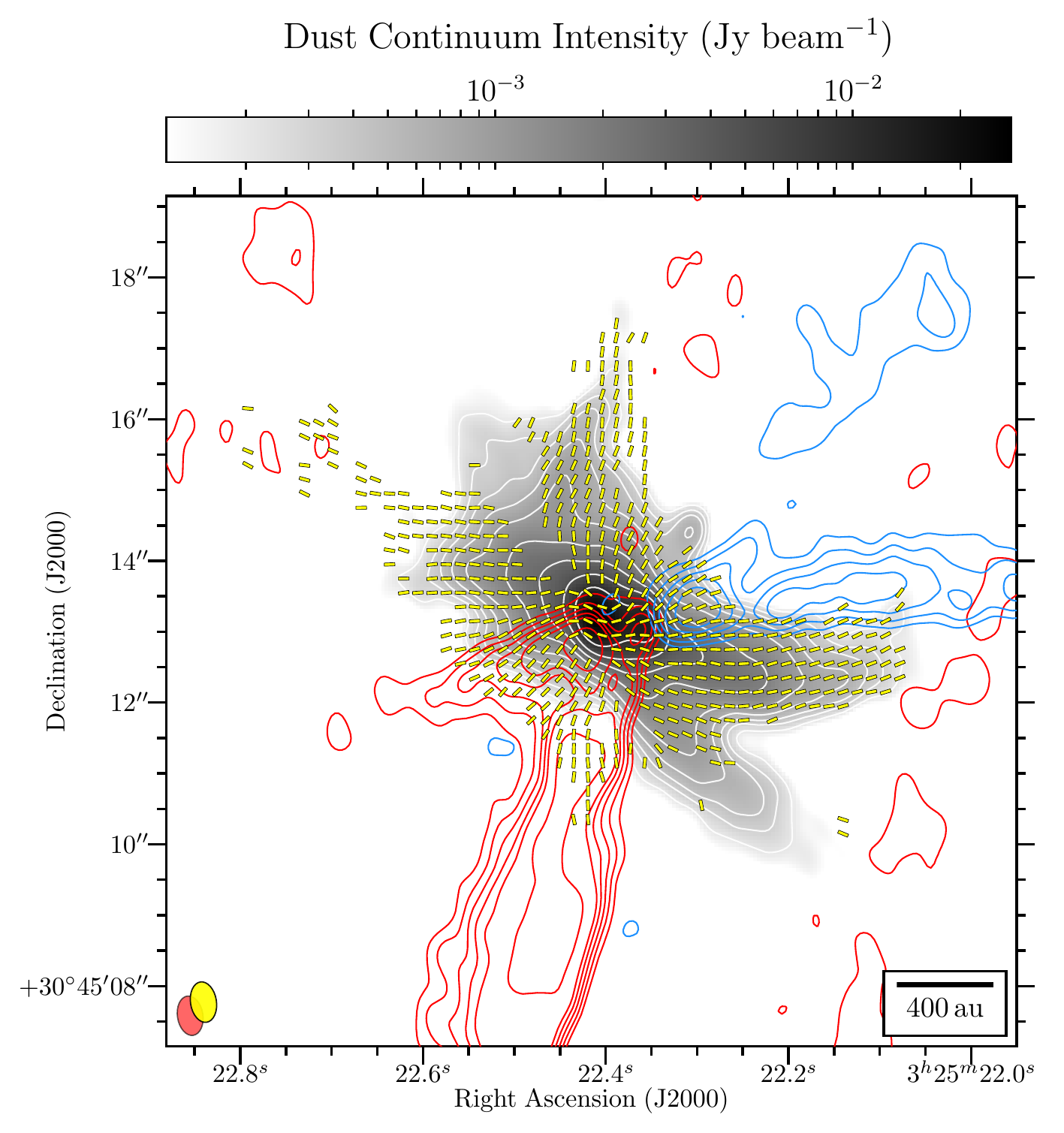}}
\subfigure{
\includegraphics[scale=0.618,clip,trim= 2cm 2cm 4cm 2cm]{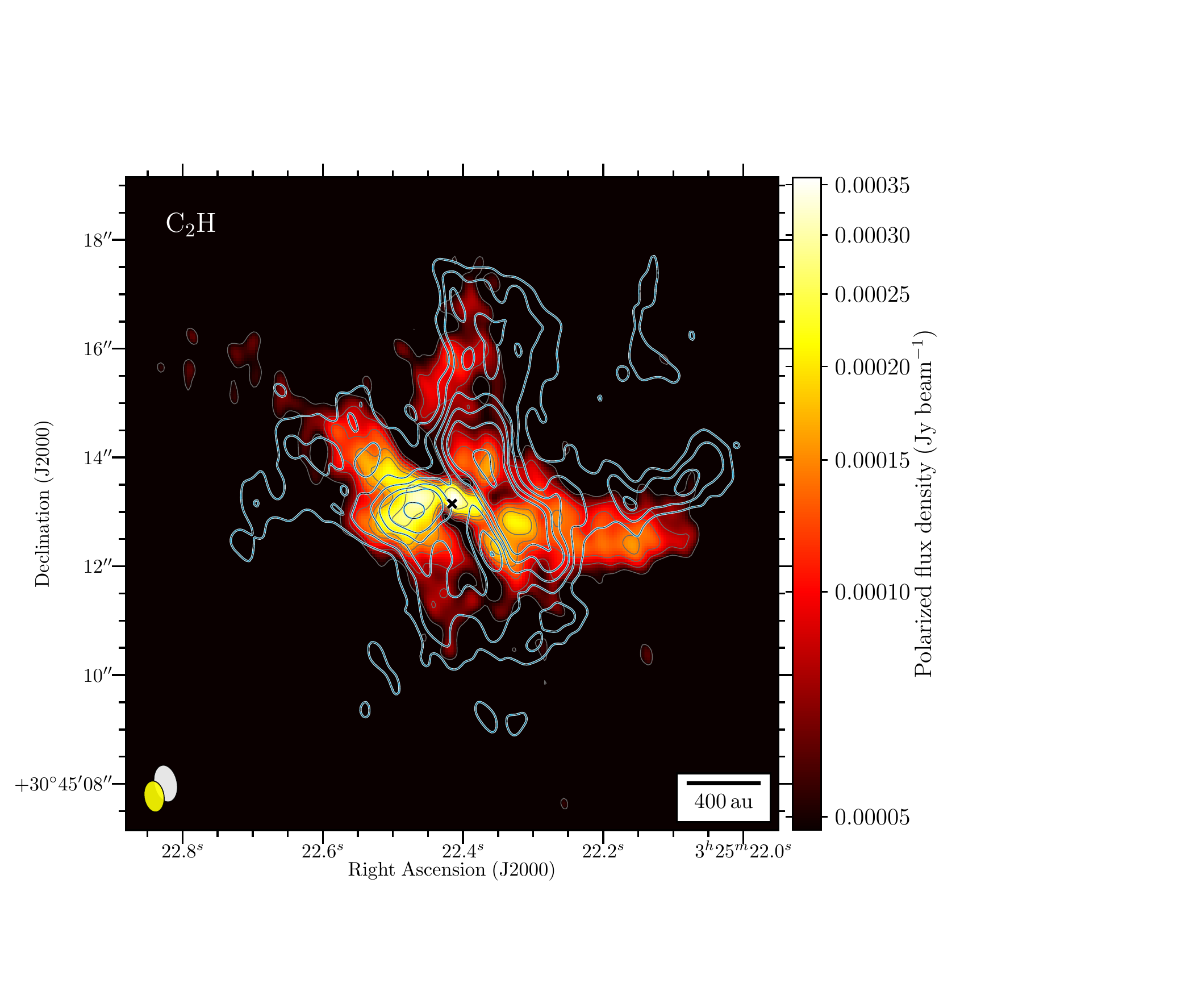}}
\caption[Magnetic fields, polarized intensity, and molecular line observations around the L1448 IRS2 Class 0 protostellar core]{\small Same as Figure \ref{fig:obs_b335_chem},  but   for the L1448 IRS2 Class 0 protostellar core.  The blue and red contours trace the moment 0 map of the blueshifted (by integrating emission from –16.5 to 3 \kms) and redshifted (by integrating emission from 6 to 21.5 \kms). The thick light blue contours (top right and bottom panels) trace the moment 0 map of the C$_2$H (by integrating emission from 3 to 8 \kms) molecular emission spectral line. The ALMA 1.3mm dust polarized emission is from \citet{Kwon2019}. The ALMA \co bipolar outflow emission is from \citet{Tobin2018}. The C$_2$H emission line is from ALMA project 2016.1.01501.S, Y. Zhang (in preparation).
}
 \label{fig:obs_l1448_chem}
\end{figure*}

\begin{figure*}[!tbh]
\centering
\hspace{-0.64cm}
\subfigure{
\includegraphics[scale=0.622,clip,trim= -0.7cm 0cm 0.5cm 0cm]{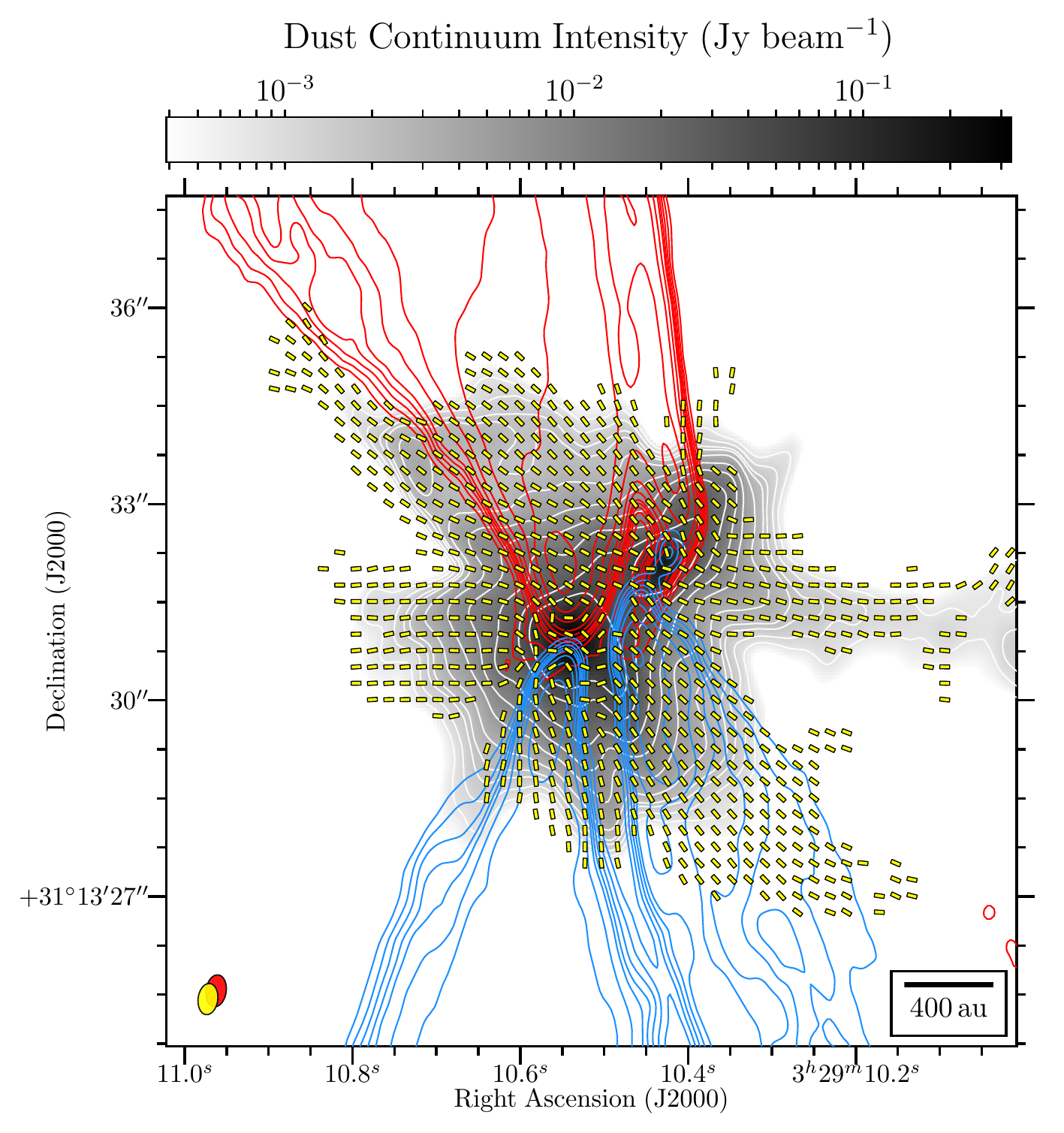}}
\subfigure{
\includegraphics[scale=0.615,clip,trim= 2cm 2cm 4cm 2cm]{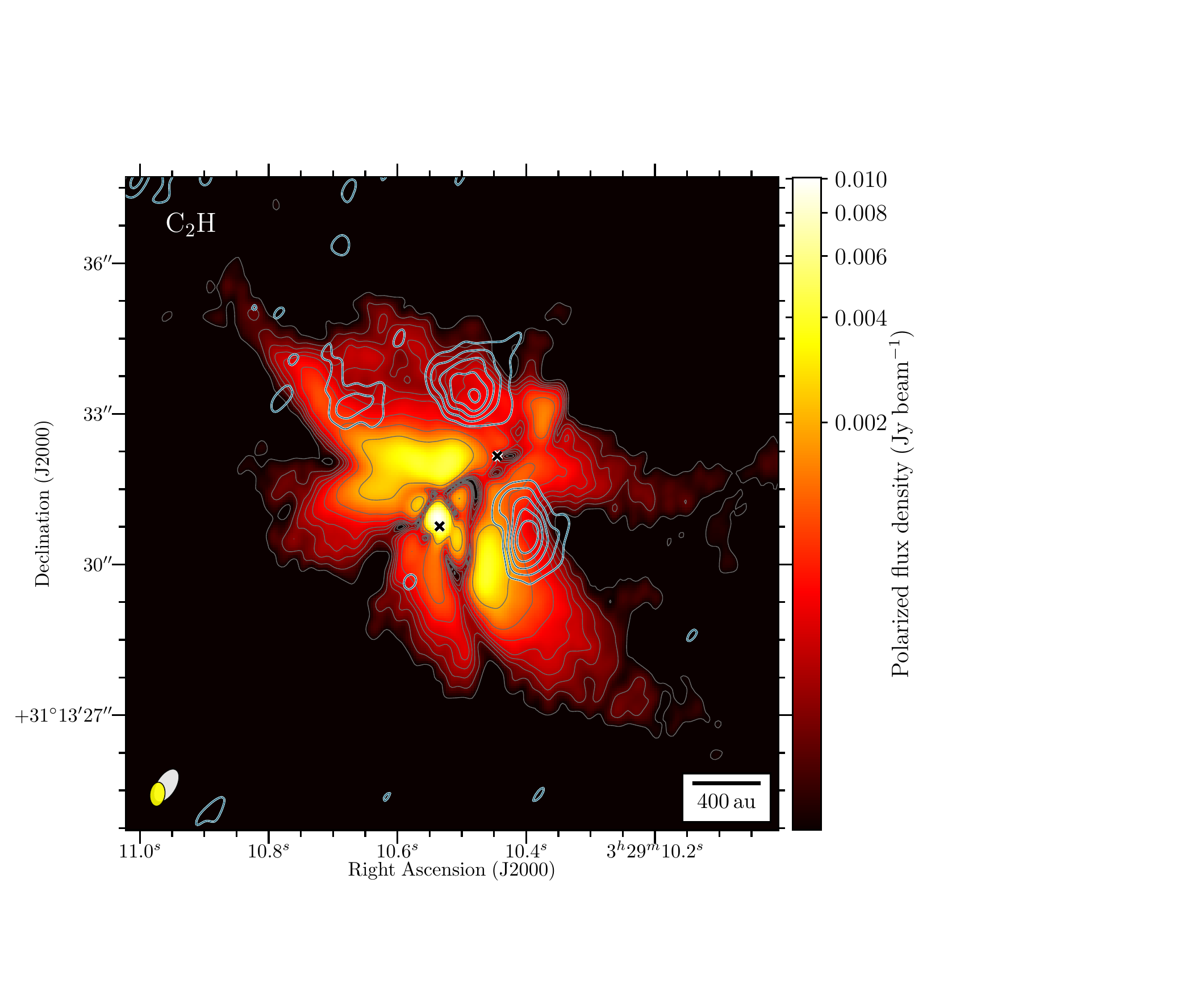}}
\\
\subfigure{
\includegraphics[scale=0.615,clip,trim= 1cm 1cm 4cm 3cm]{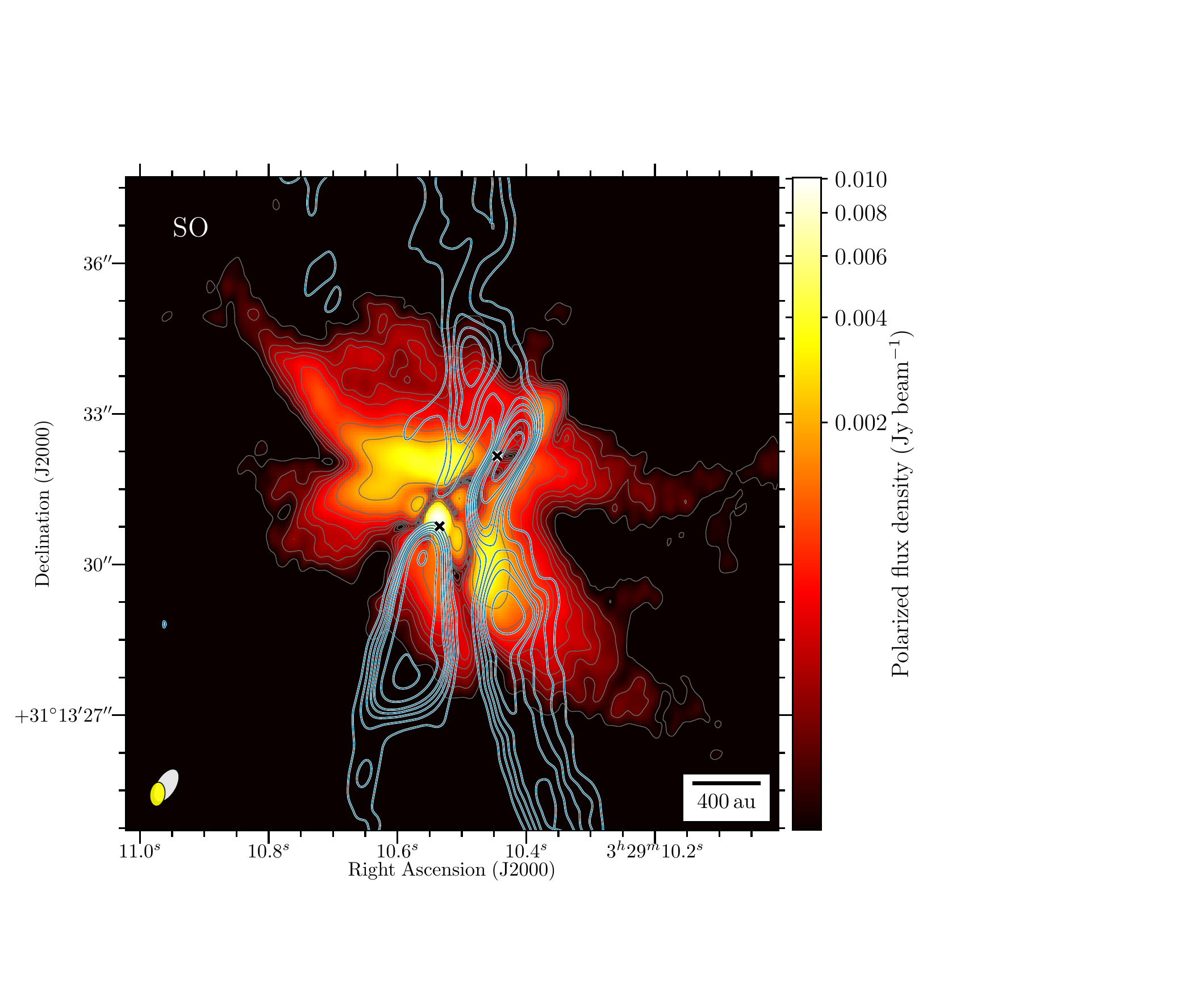}}
\caption[Magnetic fields, polarized intensity, and molecular line observations around the NC1333 IRAS4A Class 0 protostellar core]{\small Same as Figure \ref{fig:obs_b335_chem}, but for  the NC1333 IRAS4A Class 0 protostellar core.  The blue and red contours trace the moment 0 map of the blueshifted (by integrating emission from –30 to 4 \kms) and redshifted (by integrating emission from 10 to 55.5 \kms). The thick light blue contours (top right and bottom panels) trace the moment 0 map of the C$_2$H (by integrating emission from 2.5 to 15.5 \kms) and SO (by integrating emission from -10 to 19.5 \kms) molecular emission spectral line. The two crosses are located at the two peaks of the dust continuum map. IRAS4A1 is to the southeast of IRAS4A2. The ALMA 1.3mm dust polarized emission is from \citet{Ko2019}. The ALMA \co bipolar outflow emission is from T.C. Ching 2016.1.01089.S. The SO emission is line from \citet{Chuang2021}.
}
 \label{fig:obs_iras4a_chem}
\end{figure*}

\begin{figure*}[!tbh]
\centering
\hspace{-0.7cm}
\subfigure{
\includegraphics[scale=0.625,clip,trim= -0.7cm 0cm 0.5cm 0cm]{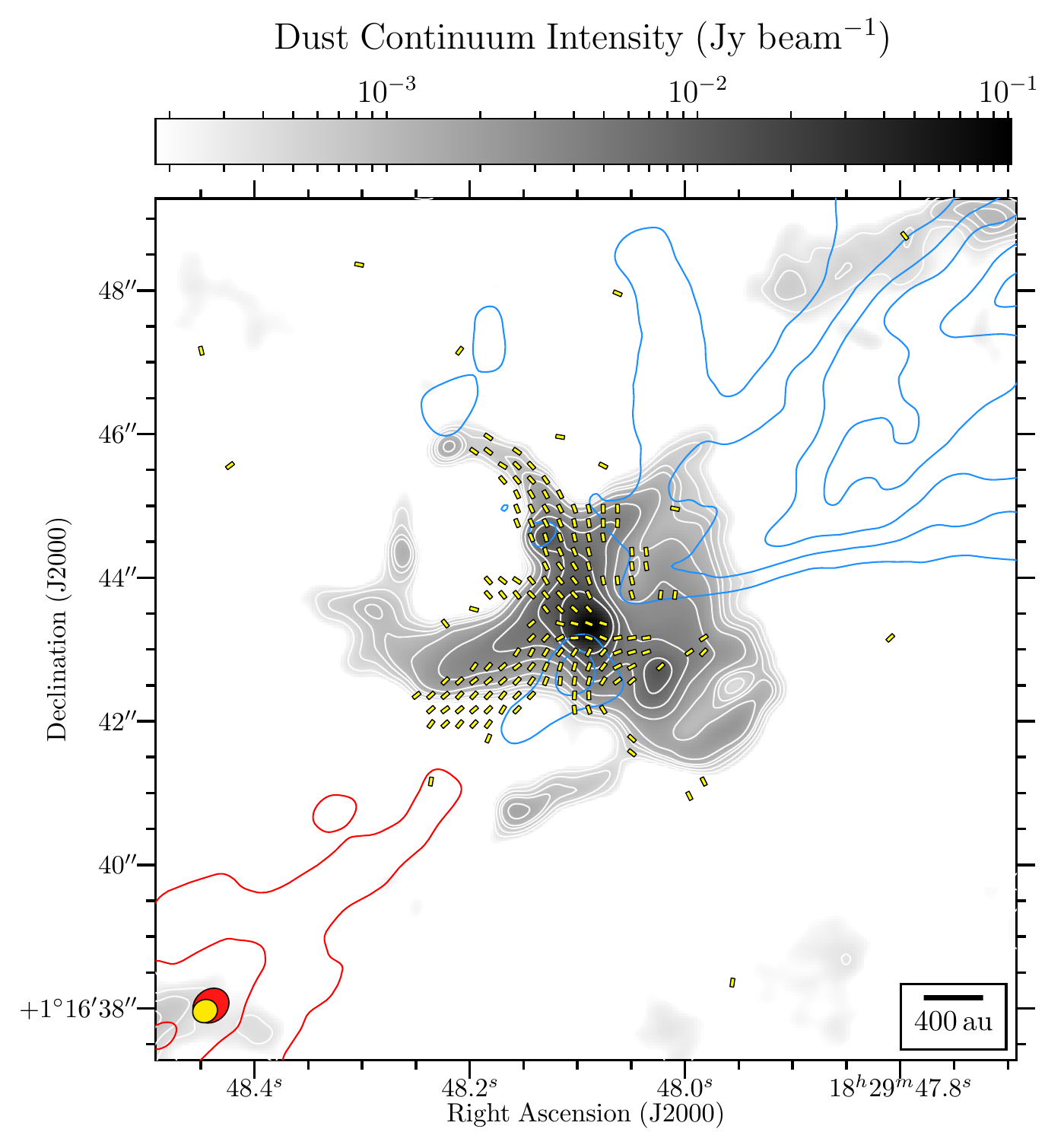}}
\subfigure{
\includegraphics[scale=0.618,clip,trim= 2cm 2cm 4cm 2cm]{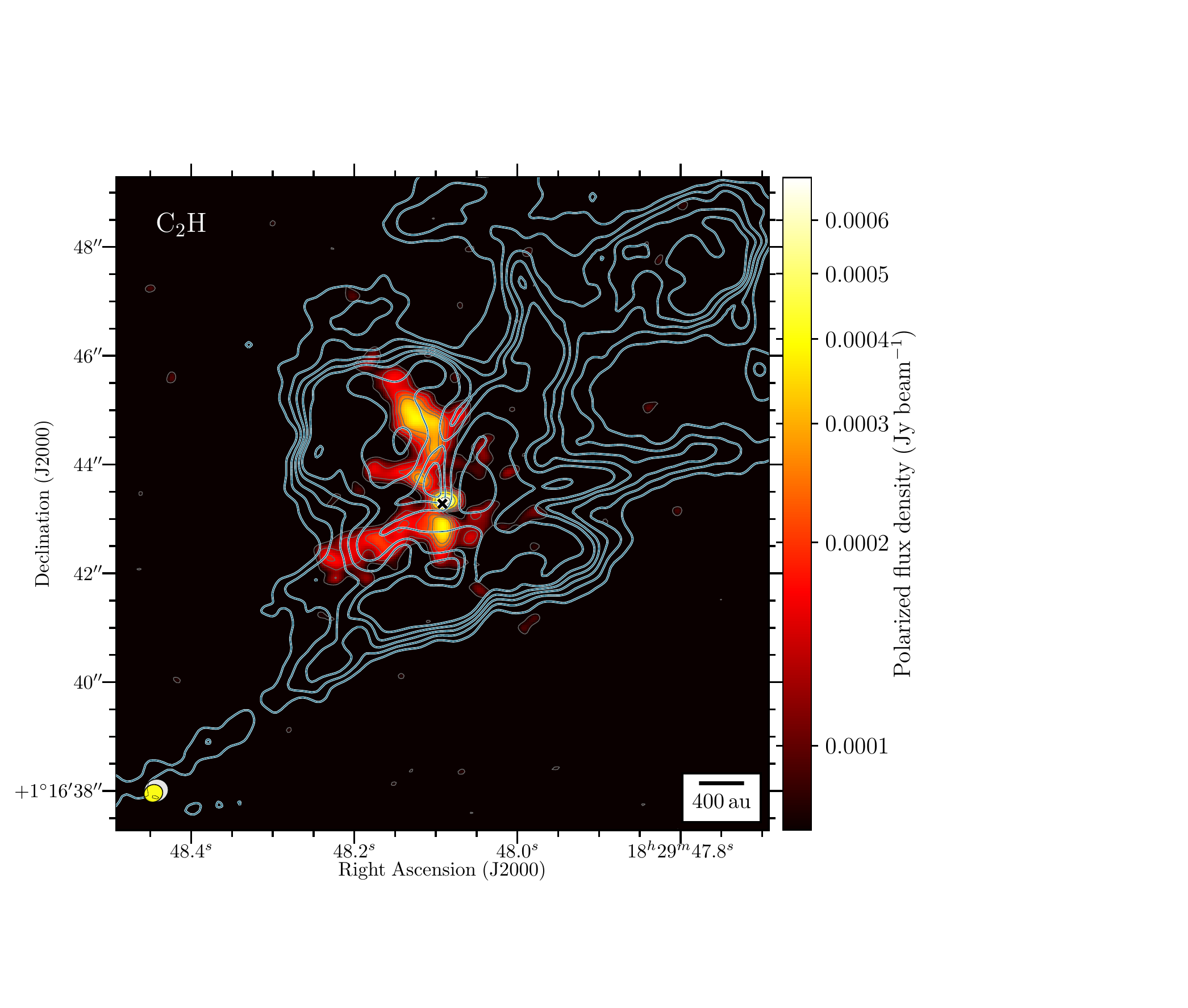}}
\\
\subfigure{
\includegraphics[scale=0.618,clip,trim= 0cm 2cm 4cm 3cm]{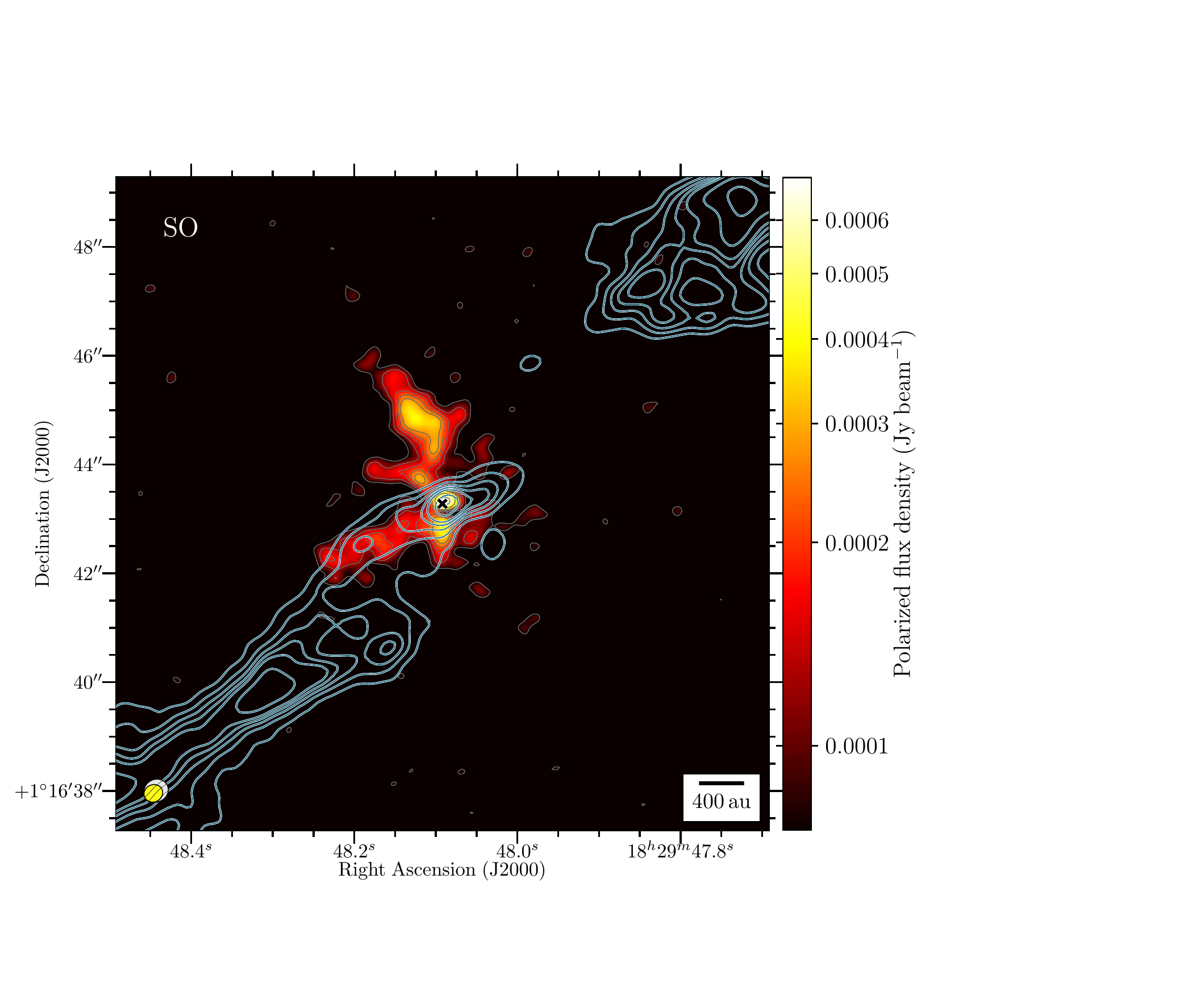}}
\caption[Magnetic fields, polarized intensity, and molecular line observations around the Serpens Emb~8 Class 0 protostellar core]{\small Same as Figure \ref{fig:obs_b335_chem}, but for  the Serpens Emb 8 Class 0 protostellar core. The blue and red contours trace the moment 0 map of the blueshifted (by integrating emission from –10 to 6.5 \kms) and redshifted (by integrating emission from 10 to 21 \kms). The thick light blue contours (top right and bottom panels) trace the moment 0 map of the C$_2$H (by integrating emission from 4.4 to 12.6 \kms) and  SO (by integrating emission from -1.5 to 15.5 \kms) molecular emission spectral line. ALMA band 0.87mm dust polarized emission is  from \citet{Hull2017b}. ALMA \co bipolar outflow emission is  from \citet{LeGouellec2019a}. C$_2$H and  SO emission line is from \citet{vanGelder2020} and \citet{Tychoniec2021}.
}
 \label{fig:obs_emb8_chem}
\end{figure*}

\begin{figure*}[!tbh]
\centering
\hspace{-0.7cm}
\subfigure{
\includegraphics[scale=0.625,clip,trim= -0.7cm 0cm 0.5cm 0cm]{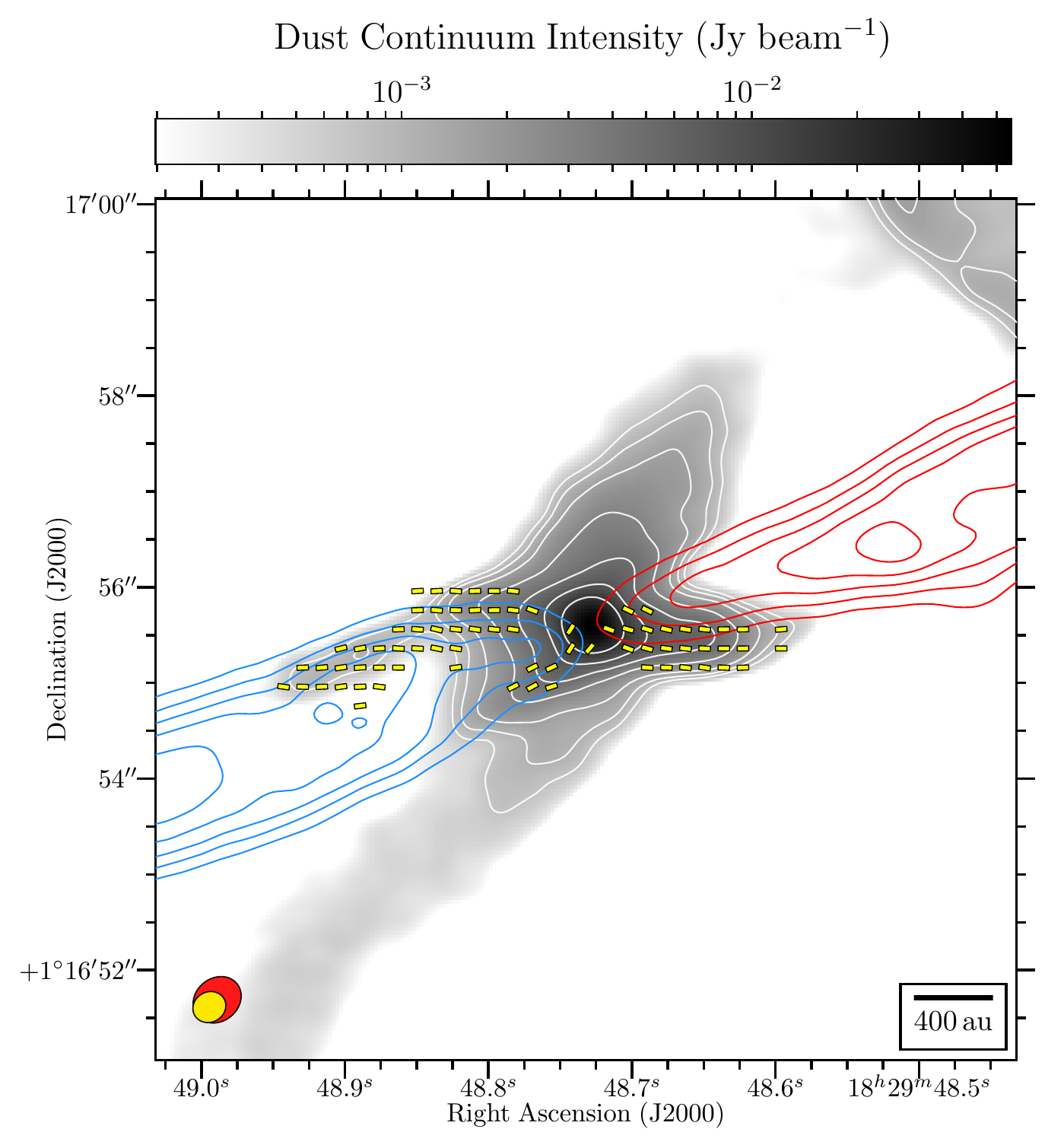}}
\subfigure{
\includegraphics[scale=0.618,clip,trim= 2cm 2cm 4cm 2cm]{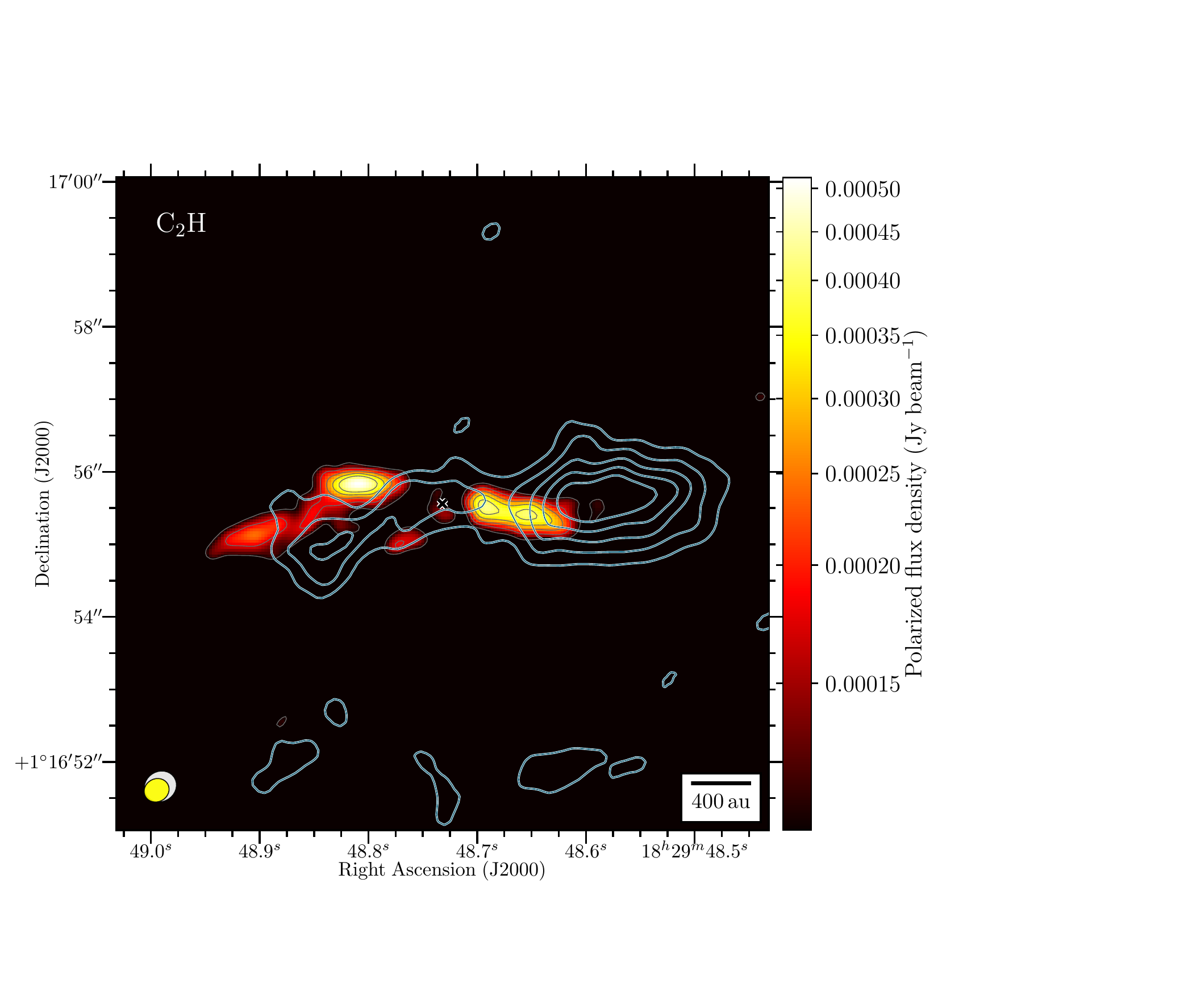}}
\\
\subfigure{
\includegraphics[scale=0.618,clip,trim= 0cm 2cm 4cm 3cm]{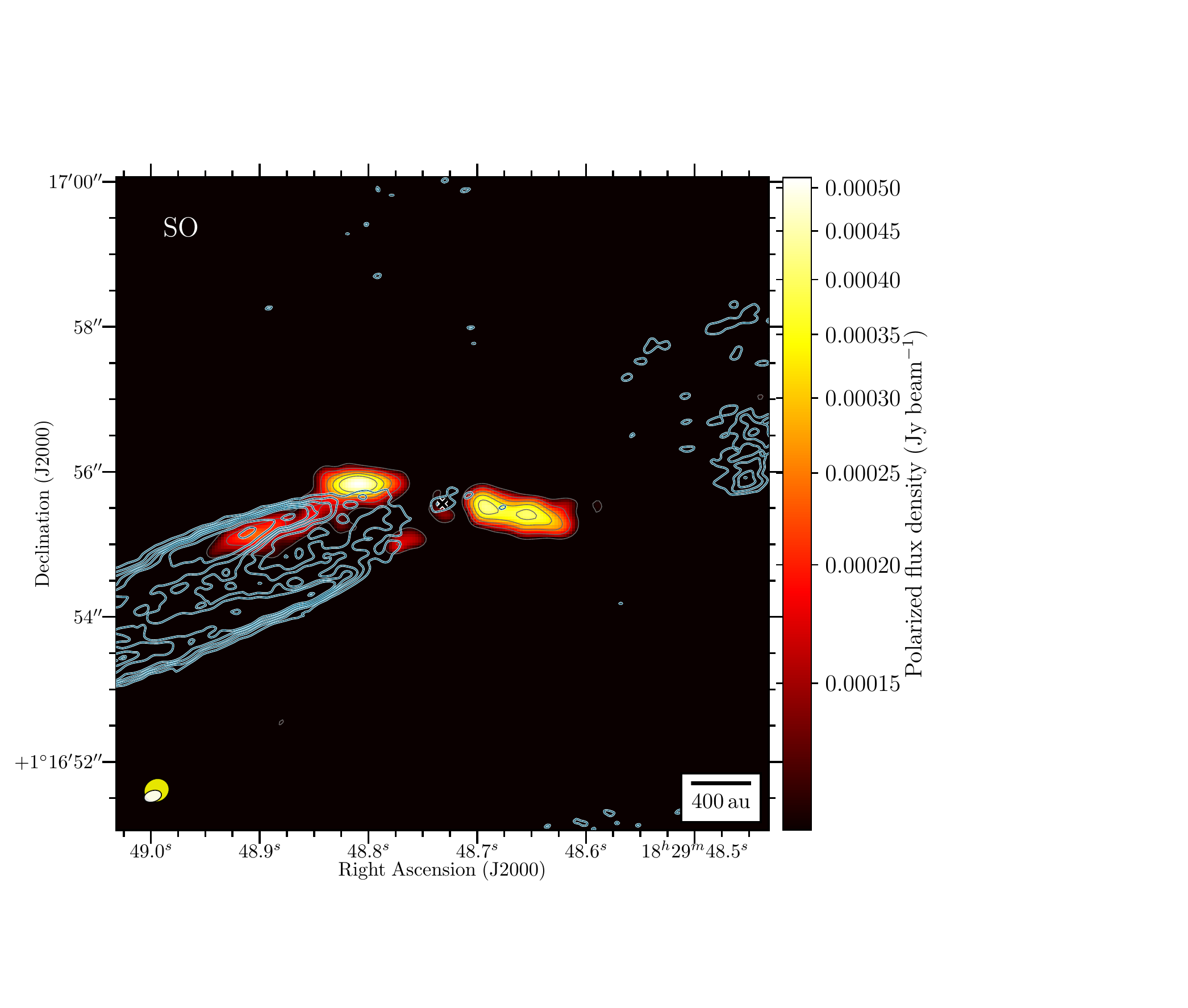}}
\caption[Magnetic fields, polarized intensity, and molecular line observations around the Serpens Emb~8(N) Class 0 protostellar core]{\small Same as Figure \ref{fig:obs_b335_chem}, but for   the Serpens Emb 8(N) Class 0 protostellar core.   The blue and red contours trace the moment 0 map of the blueshifted (by integrating emission from –10 to 6.5 \kms) and redshifted (by integrating emission from 10 to 21 \kms). The thick light blue contours (top right and bottom panels) trace the moment 0 map of the C$_2$H (by integrating emission from 5.4 to 12.8 \kms) and  SO (by integrating emission from -16 to 8 \kms, and from 10.5 to 35.5 \kms) molecular emission spectral line. The ALMA band 0.87mm dust polarized emission, ALMA \co bipolar outflow emission, and C$_2$H emission line are from \citet{LeGouellec2019a}. The SO emission line is from  ALMA project 2019.1.00931.S, Tychoniec et al. (in prep.).
}
 \label{fig:obs_emb8N_chem}
\end{figure*}

We collected and analyzed observations of six Class 0 protostars carried out with the ALMA interferometer: B335, L1448 IRS2, NGC1333 IRAS4A1 \& IRAS4A2, Serpens Emb 8, Serpens Emb 8(N). We present in  Figures \ref{fig:obs_b335_chem}, \ref{fig:obs_l1448_chem}, \ref{fig:obs_iras4a_chem}, \ref{fig:obs_emb8_chem}, and \ref{fig:obs_emb8N_chem} the maps of the dust continuum emission, the polarized dust emission, and the molecular gas emission obtained toward six sources. We use molecular line observations to either trace the bipolar outflow (with the \co transition), or to probe the irradiation field in the inner core with the C$_2$H and SO molecular species (two other molecules $c$-C$_3$H$_2$ and CS, potentially tracing the radiation field, are presented in Appendix \ref{sec:app_other_lines} as additional material). As introduced above, the C$_2$H molecule has been detected in PDRs, and the role of the irradiation field (enhancing the ionization) in its chemical formation pathway has been well established. In addition, SO is a tracer of shocks, which can be sources of UV  photons.  We compare the maps of polarized flux with those of  moment 0 molecular emission lines to investigate whether the dust polarization is enhanced in highly irradiated zones. The goal is to understand the physical conditions that favor efficient grain alignment. We  discuss the chemistry of these molecules in Section \ref{sec:disc_UV}.\\

Details of the sources and dust polarization observations we present can be found in Table \ref{t.obs_pol}, while the details of the spectral lines studied can be found in Table \ref{t.obs_lines}. These sources were selected because  the interferometric maps exhibit extended polarized dust emission throughout the inner envelope ($\sim$10-2000\,au), and ALMA datasets of the chemical tracers we target were publicly available, offering similar $uv$-coverage to the dust polarization observations\footnote{Another source, the Class 0 B1-c in Perseus, originally triggered our interest. The ALMA C$_2$H molecular emission line published in \citet{vanGelder2020,Tychoniec2021} exhibit an extended emission toward the bipolar outflow cavity and cavity walls. However, the 1.3 mm ALMA dust polarization observations published in \citet{Cox2018} suffers from a lack of sensitivity; only a few patches of polarized intensity are detected at the base of the outflow cavity.}. The polarized dust emission maps of B335, Serpens Emb 8, and Serpens Emb 8(N) are  from \citet{Hull2017a}, \citet{LeGouellec2019a}, and \citet{LeGouellec2020}, and details of the data reduction can be found in the corresponding references. For this project we   calibrated and imaged the ALMA dust polarization observations toward L1448 IRS2, and NGC1333 IRAS4A. To produce the polarized dust continuum emission maps we used the \texttt{tclean} task of the CASA software version 5.8 \citep{McMullin2007}. We applied three to four rounds of consecutive phase-only self-calibration \citep{Brogan2018}, using the total intensity (Stokes I) solutions as the model, with a Briggs weighting parameter of 0.5 to 1, depending on the source, and the appropriate integration time intervals. The three Stokes parameters $I$, $Q$, and $U$ were cleaned separately after the last round of self-calibration, using an appropriate residual threshold and number of iterations. We then created the debiased polarized intensity maps following the method from \citet{WardleKronberg1974} and \citet{Hull2015b}. 

A variety of ALMA 12m datasets at 1.3 mm were used to target the molecular emission lines of C$_2$H and SO, whose angular resolutions are about 0.3$\farcs$--0.8$\farcs$. The datasets were pipeline calibrated and manually phase-only self-calibrated\footnote{For the self-calibration of the spectral lines, we apply the phase self-calibration solutions calculated with the continuum maps devoid of spectral lines on the individual spectral line measurement sets.}, with the CASA software version 5.8 \citep{McMullin2007}. The images were made with the \texttt{tclean} task. The images were produced with an adapted mask constructed manually in each channel covering the entire emission. For the outflow traced by the CO molecular line, we pipeline calibrated and manually phase-only self-calibrated ALMA 12m observations of the \co transition for L1448 IRS2 and NGC1333 IRAS4A. The CO molecular line observations are from \citet{Cabedo2021} in B335, \citet{Hull2017a} in Serpens Emb 8, \citet{LeGouellec2019a} and \citet{Tychoniec2019} in Serpens Emb 8(N), and \citet{Tobin2018} in L1448 IRS2.

\textbf{B335:} Figure \ref{fig:obs_b335_chem} presents the dust polarization observations 
 from \citet{Maury2018} and Maury et al. (in preparation) overlaid with contours following the intensity levels of the integrated emission of the C$_2$H molecular line from \citet{Imai2016}. The dataset also includes the SO ($N_J$=6$_7 \rightarrow 5_6$) transition, which   we do not show here as the line is only detected at the dust continuum peak. As pointed out in \citet{Maury2018}, the polarized dust emission in B335 is bimodal. The polarized flux is either enhanced along the outflow cavity walls with   the average magnetic fields orientated east--west following the walls of the cavities or enhanced toward the equatorial plane of the core (the northern patch of polarized dust emission) where the magnetic field orientations are north--south. This bi-modality is also found in the histogram of polarization position \citep{Maury2018}, and is actually present in a few other published observations of Class 0 cores \citep{LeGouellec2020}. The emission of the carbon chain molecule C$_2$H is clearly enhanced toward the outflow cavities. C$_2$H emission appears across the bulk of the bipolar cavity traced by the CO emission, with enhancement along the cavity walls, up to $\sim$1500 au for the southern edges. The polarized flux and these molecular emission lines thus appear to spatially overlap toward the outflow cavity walls. However, the equatorial plane to the north of the dust continuum peak, which is strongly polarized, seems to be devoid of molecular line emission tracing irradiated conditions, suggesting a different nature of the irradiation conditions in the mid-plane, compared to the outflow cavity walls. This applies only to the northern part of the core, as the southern part is much less polarized.\\

\textbf{L1448 IRS2:} Figure \ref{fig:obs_l1448_chem} presents the dust polarization observations from \citep{Kwon2019} overlaid with contours following the intensity levels of the integrated emission of the C$_2$H molecular line from Y. Zhang et al. (in preparation) (ALMA project 2016.1.01501.S, PI: N. Sakai). The dataset also includes the SO ($N_J$=6$_7 \rightarrow 5_6$) and SO ($N_J$=6$_6 \rightarrow 5_5$) transitions, which  we do not show here as the emission only peaks at the dust continuum peak. The polarized dust emission is enhanced on each side of blue- and redshifted outflow, embracing the cavities. The magnetic field lines exhibit a very pinched hourglass-like morphology. The depolarization in the mid-plane is likely due to beam smearing effects. The C$_2$H molecular emission line appears enhanced toward the walls of the cavities, as is the case for the polarized dust emission.\\

\textbf{NGC1333 IRAS4A:} Figure \ref{fig:obs_iras4a_chem} presents the dust polarization observations from \citet{Ko2019} overlaid with contours following the intensity levels of the integrated emission of the C$_2$H and SO molecular lines from \citet{Chuang2021}.  Similarly to the SMA dust polarization of this wide binary \citep{Girart2006}, the magnetic field global morphology exhibits a pinched hourglass-like morphology aligned with the direction of the bipolar outflows. The maps shown here, however, suggest that the polarized emission is enhanced along the outflow cavities of both protostars IRAS4A1 (southeast) and IRAS4A2 (northwest) as well as the infalling equatorial planes. The SO molecular line peaks toward the blue- and redshifted outflow of IRAS4A2, and only in the blueshifted lobe of the IRAS4A1 outflow. In such a bright Class 0 object, where dust polarization is strong and extended across the inner $\sim$2000 au of the core, C$_2$H emission is surprisingly not extended compared to the first two   objects we described. The C$_2$H molecular emission line is mostly associated with two compact emission features in the outflow of IRAS4A2, and a weak $\sim$5$\sigma$ detection toward the eastern wall of the IRAS4A1 redshifted cavity. The lower S/N emission shown in \citet{Chuang2021} exhibits a more extended emitting region, consistent with the outflow cavity.\\

\textbf{Serpens Emb 8:} Figure \ref{fig:obs_emb8_chem} presents the dust polarization observations from \cite{Hull2017b} overlaid with contours following the intensity levels of the integrated emission of the C$_2$H and SO molecular lines from \citet{vanGelder2020}. In this core the scenario seems to be   surprising, given the drastic difference between the extent of the C$_2$H emission with respect to the polarized dust emission. The C$_2$H emission  extends much farther than the polarized dust emission in the central regions of the envelope. Where it overlaps with the polarized dust emission, the C$_2$H emission morphology seems to show a morphology following closely that of the polarized intensity, especially to the  southeast of the core. At the base of the cavity walls of the blueshifted outflow, the C$_2$H emission is also enhanced and exhibits a V-shaped pattern, confirming that the cavity walls are the ideal conditions to form and/or excite the rotational transitions of C$_2$H. The SO and C$_2$H peaks are  also $\sim$ 2000 au away from the center, toward the cavity of the blueshifted outflow in the northwest, where faint dust emission is also detected. 
The surprising morphology of the C$_2$H emission in this source, which is more extended than the emission in the other protostars analyzed here, could be explained by the recent accretion activity of this source. With near-infrared spectroscopy, \citet{Greene2018} showed that the properties of the central protostellar embryo (\ie low surface gravity and effective temperature) suggest that Serpens Emb 8 has recently undergone a period of high accretion activity, which may explain the spatial extent of the irradiation-sensitive chemistry in the core.
The emission from the SO line exhibits an elongated structure toward the center of the core that overlaps with one patch of polarized dust emission. Finally,   SO   also peaks on the southern side of the redshifted outflow lobe. \\

\textbf{Serpens Emb 8(N):}  Figure \ref{fig:obs_emb8N_chem} presents the dust polarization observations from \citet{LeGouellec2019a} overlaid with contours following the intensity levels of the integrated emission of the C$_2$H and SO molecular lines from \citet{LeGouellec2019a} and Tychoniec et al. (in preparation). C$_2$H peaks in the outflow cavities, with an asymmetry in the emission strength (it is stronger on the redshifted side); it is slightly spatially shifted from the polarized dust emission. On the redshifted side C$_2$H peaks farther away than the polarized dust emission, and on the blueshifted side C$_2$H peaks inside the cavity, while the polarized dust emission is enhanced toward the walls of the cavity. The disorganization of the structures in dust and polarized dust emission of Serpens Emb 8, relative to Serpens Emb 8(N), which is much more symmetric and organized, suggests that irradiation may have escaped more easily throughout the inner core in Emb 8. This would explain why the C$_2$H emission is so different between the two sources: it only peaks in the walls of the cavity in Emb~8(N) versus everywhere in the inner envelope and in the blueshifted outflow lobe in Emb 8.
In Serpens Emb 8(N), the SO molecular emission line peaks toward the walls of the blueshifted cavity, from the center of the source up to $\sim$ 2000 au. It also peaks in the redshifted cavity, but much farther away, at $\sim$ 2000 au from the source, likely spotting the bow shock studied in \citet{Tychoniec2019}. Toward the northern wall of the blueshifted cavity, the SO emission coincides with a patch of highly polarized dust emission.
\\

\begin{figure*}[!tbh]
\centering
\subfigure{
\includegraphics[scale=0.3,clip,trim= 0.5cm 1cm 2cm 2cm]{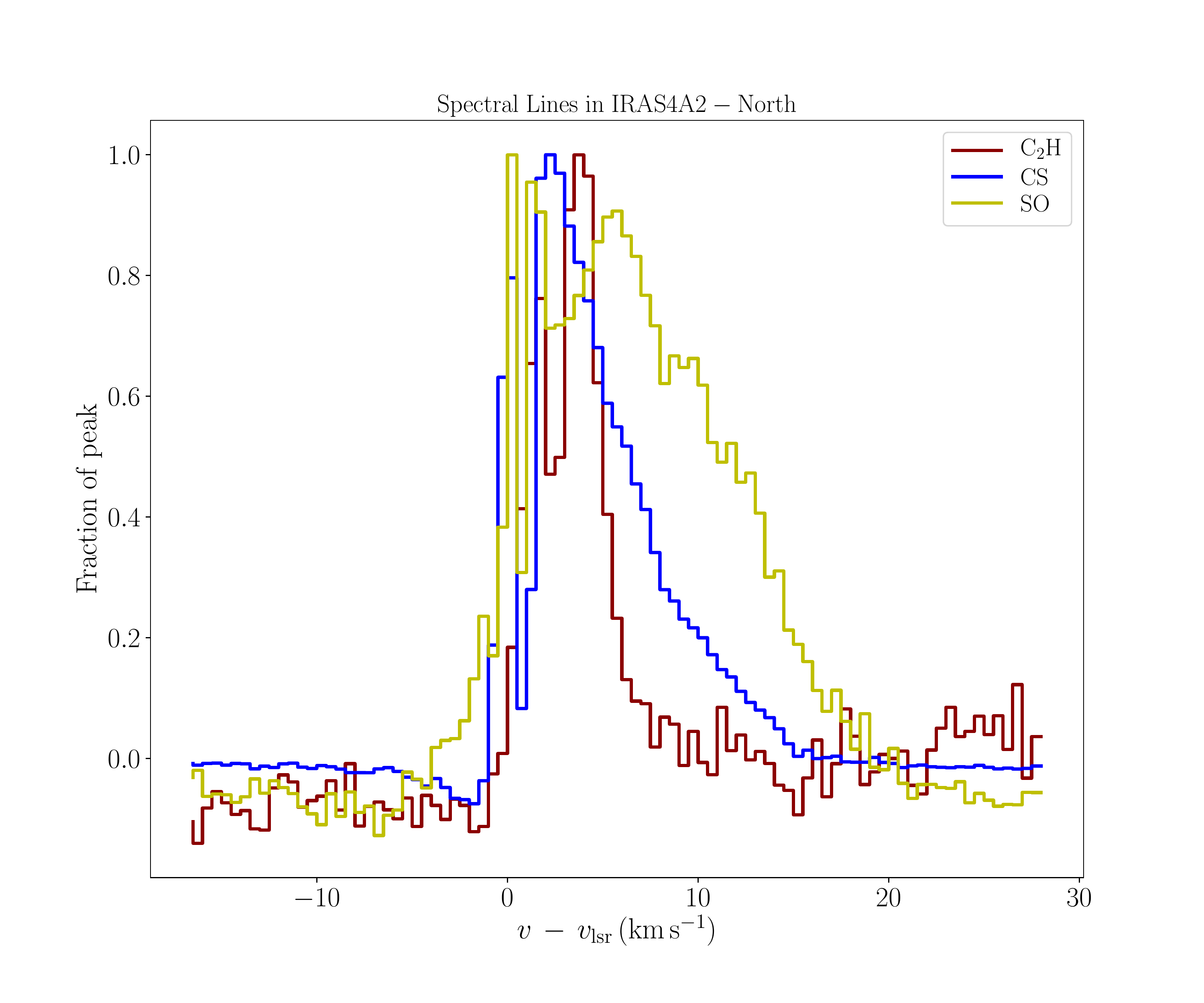}}
\subfigure{
\includegraphics[scale=0.3,clip,trim= 0.5cm 1cm 2cm 2cm]{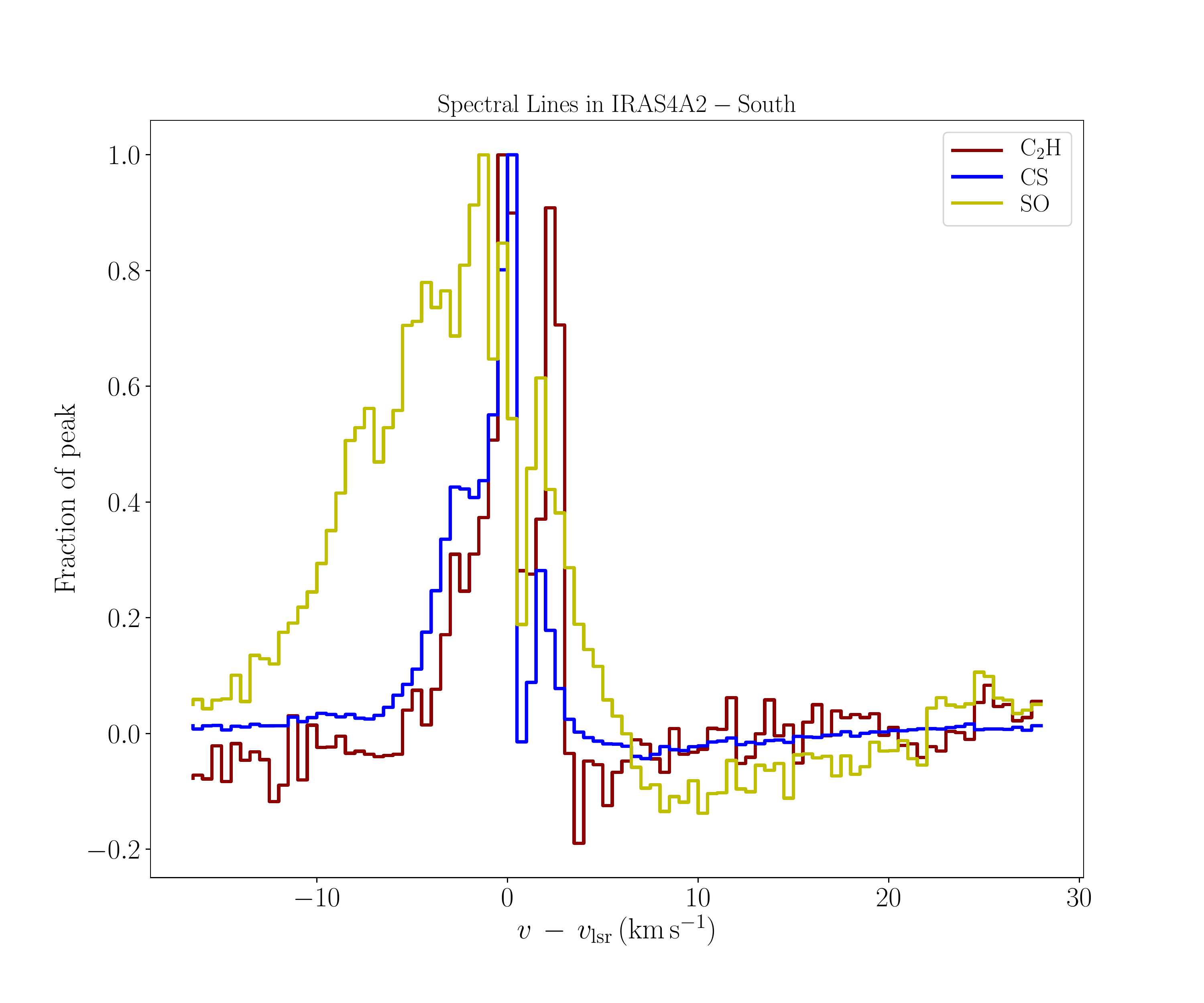}}
\\
\subfigure{
\includegraphics[scale=0.3,clip,trim= 0.5cm 1cm 2cm 2cm]{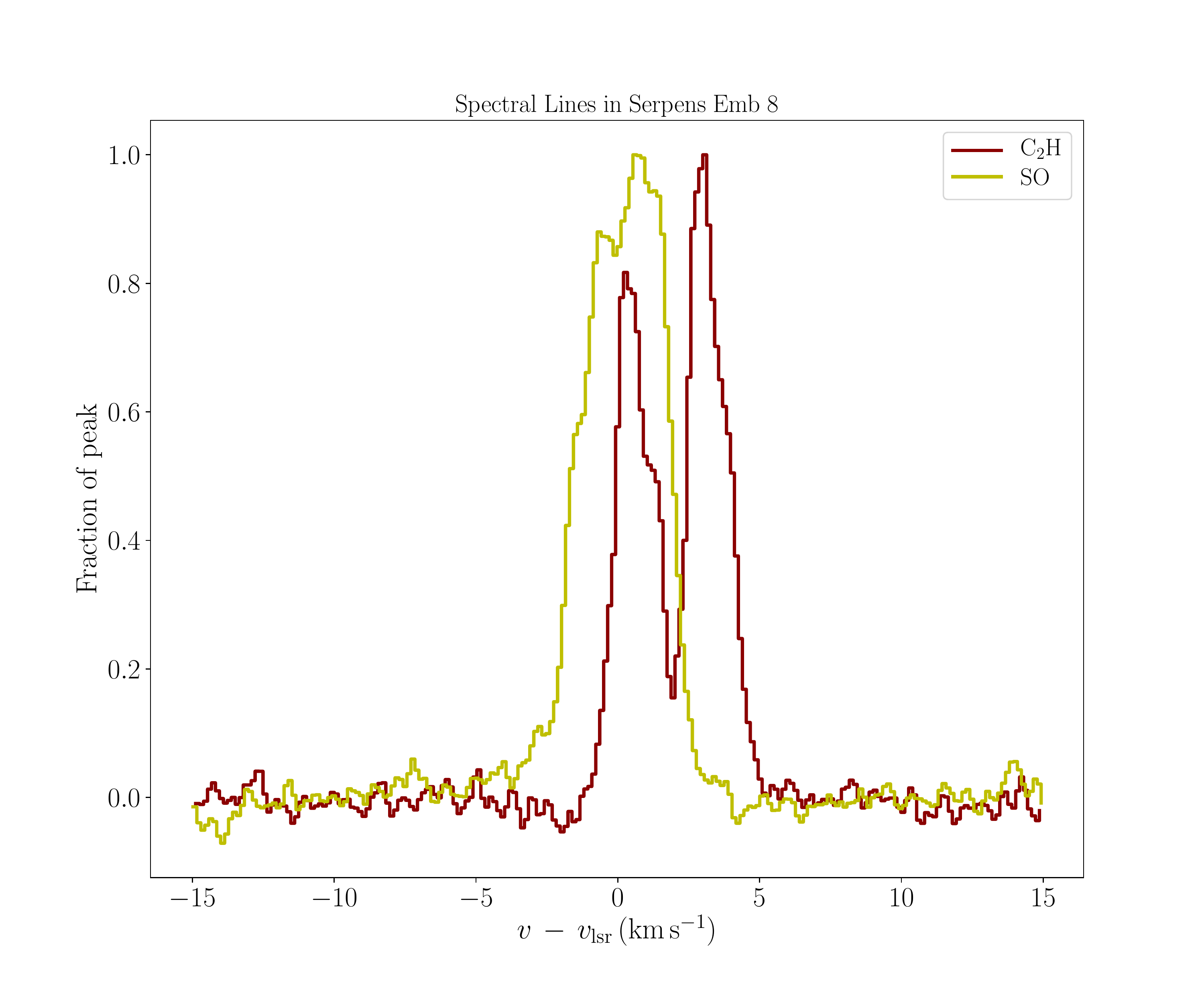}}
\subfigure{
\includegraphics[scale=0.3,clip,trim= 0.5cm 1cm 2cm 2cm]{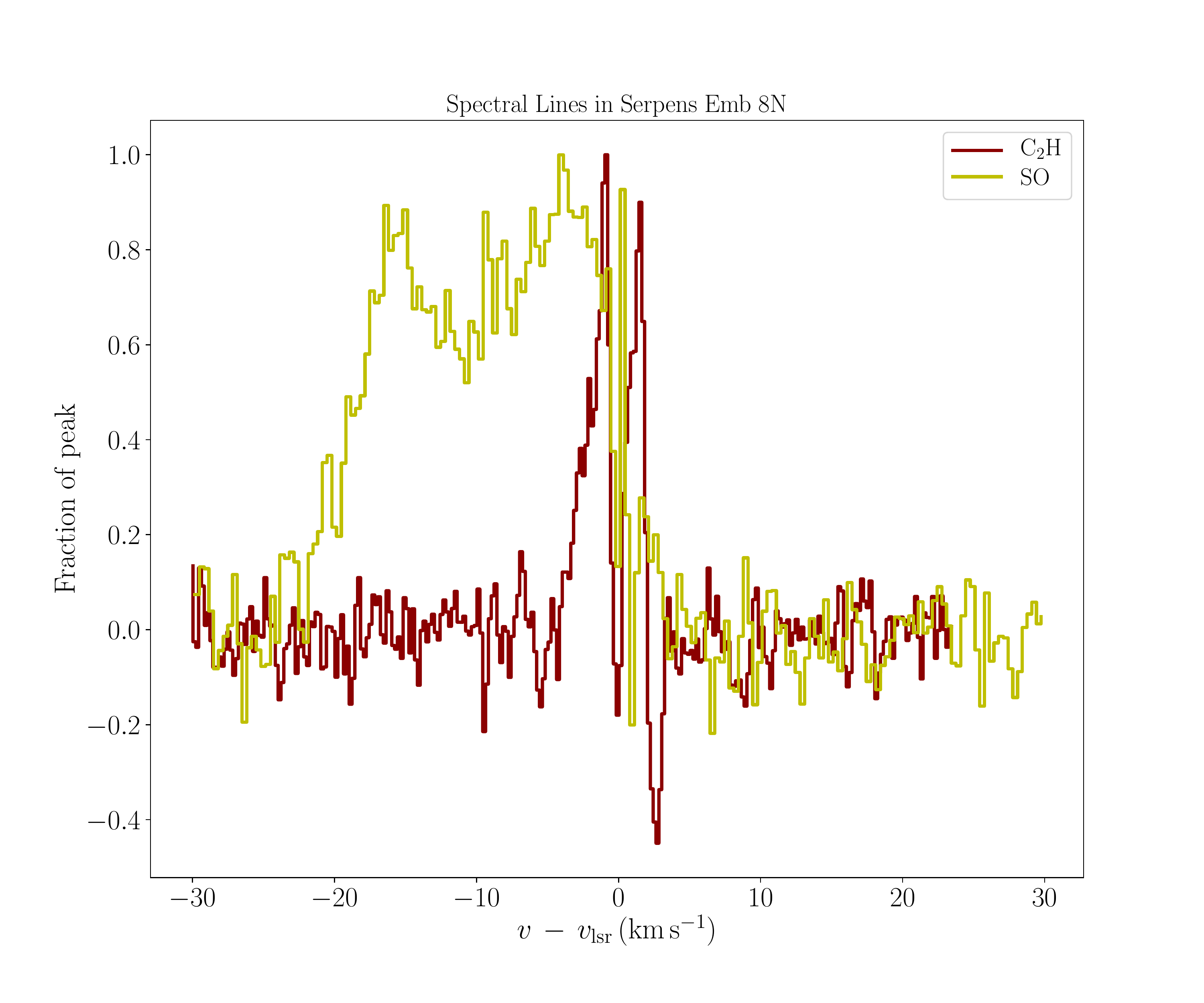}}
\caption{\small Spectra of molecular emission lines in NGC1333 IRAS4A2, Serpens Emb8, and Serpens Emb 8N. 
The solid lines show the spectra of the C$_2$H, CS, and SO molecular emission lines, toward the north (\textit{top left panel}) and the south  (\textit{top right panel}) of IRAS4A2. 
The spectra of the C$_2$H and SO molecular emission lines toward the redshifted cavity of Serpens Emb 8, and the blueshifted cavity of Serpens Emb 8(N), are shown in the \textit{bottom left} and \textit{bottom right panel}, respectively. For IRAS4A, the emission is summed over a circle of 2.5$\farcs$ in diameter, centered on the two peaks of C$_2$H visible in the integrated moment 0 map seen in Figure \ref{fig:obs_iras4a_chem}. For Serpens Emb 8 and Serpens Emb 8(N), the spectra are from a circle of 1$\farcs$ in diameter located at 800 au from the center, in the redshifted and blueshifted cavity, respectively. In all cases, while the C$_2$H hyperfine structure is detected at the $v_{\rm{lsr}}$, the spectra of CS and SO are clearly broader, suggesting material linked to the outflowing gas.}
\label{fig:obs_iras4a2_spectra}
\end{figure*}

In summary, the carbon chain molecule C$_2$H seems preferentially associated with outflows and cavity walls in some sources, and is widely distributed in others. 
Such different spatial segregation may suggest that the source geometry plays an important role in the heating of the inner envelope (see Section \ref{sec:KS_disc}). Another spatial segregation among various irradiation conditions is also suggested by the emission line spectra in IRAS4A, Serpens Emb 8, and Serpens Emb 8(N) shown in Figure \ref{fig:obs_iras4a2_spectra}.
In these four spectra, while the velocity of the C$_2$H associated hyperfine structure is distributed around the systemic velocity of the source, SO exhibits a much broader range of gas velocities. This indicates to what extent the molecular species we target belong to the outflow (or the material entrained by the outflow),  like SO, or if it belongs to the envelope--cavity walls, like C$_2$H. In addition, the spectra of CS is distributed toward a broader range of gas velocities than the C$_2$H spectra, but narrower compared to the SO spectra (see Appendix \ref{sec:app_other_lines} for the moment 0 maps of CS). Hence, while C$_2$H seems to trace the irradiated walls of the cavity, located between the outflowing gas of the outflow and the infalling gas from the envelope, SO seems to be linked with the outflowing gas, and as a shock tracer its emission emanates from the interaction zone between cavity walls and the jet--outflow system.

These molecular lines are expected to trace the regions with abundant UV photons in the gas.
We  thus investigate whether the polarized dust emission presents different properties in these areas, which would allow us to characterize the role played by such energetic photons in the mechanisms producing the polarized dust emission. However, significant dust polarization signals are observed toward locations that are likely not under a strong radiation field, for example in equatorial mid-planes and envelope emission in the north of B335 and mid-planes of IRAS4A. Therefore, the local physical conditions need to be further explored in order to understand how dust grain alignment is efficient in this kind of  environment. This is addressed with the help of radiation transfer calculations presented in Paper II.

\section{Comparing the spatial distribution of the emissions maps}
\label{sec:p3_obs_KS}

\begin{figure*}[!tbh]
\centering
\subfigure{
\includegraphics[scale=0.6,clip,trim= 2.2cm 1cm 0cm 0cm]{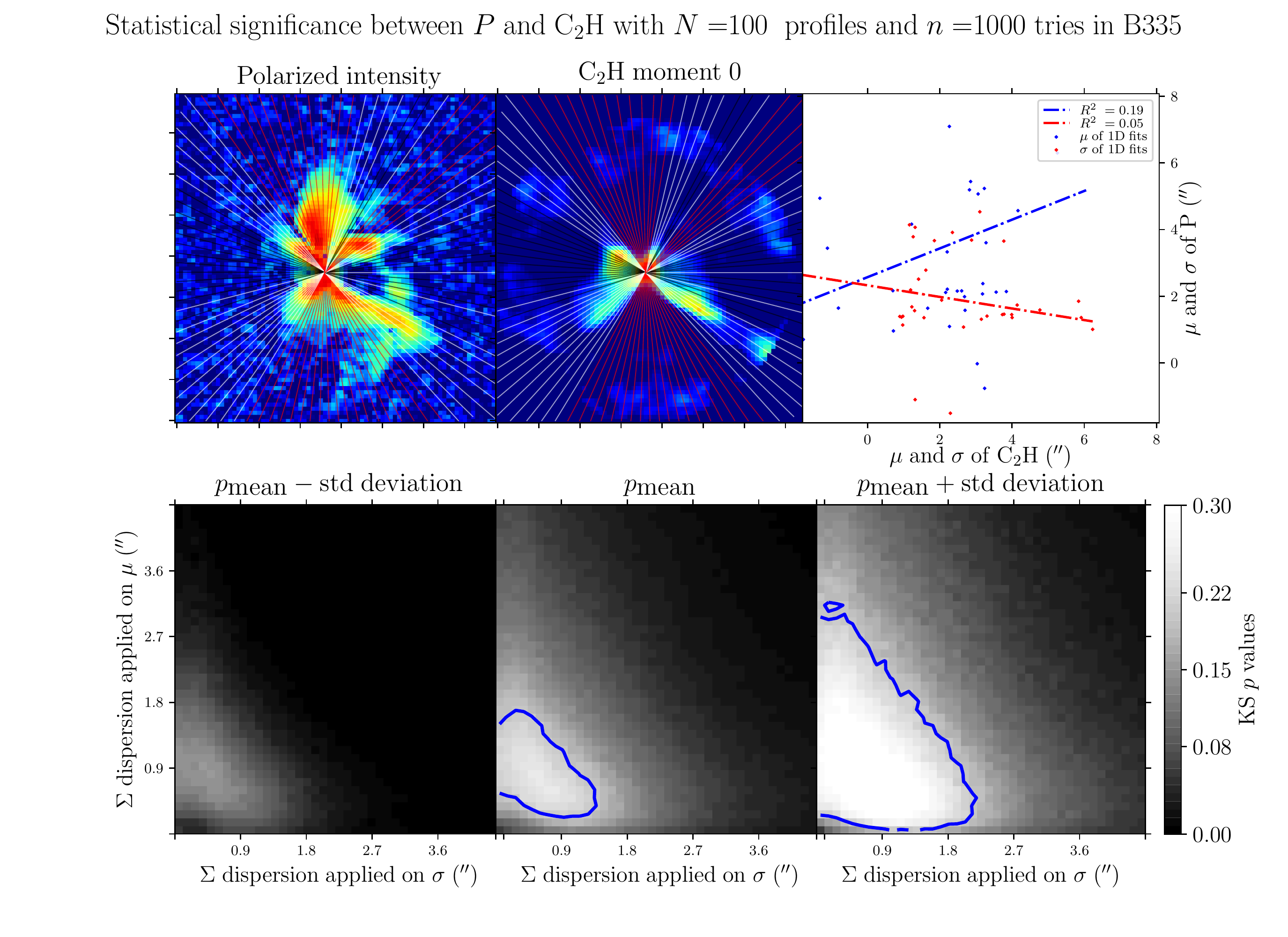}}
\subfigure{
\includegraphics[scale=0.55,clip,trim= 1.5cm 0cm 1.5cm 0.4cm]{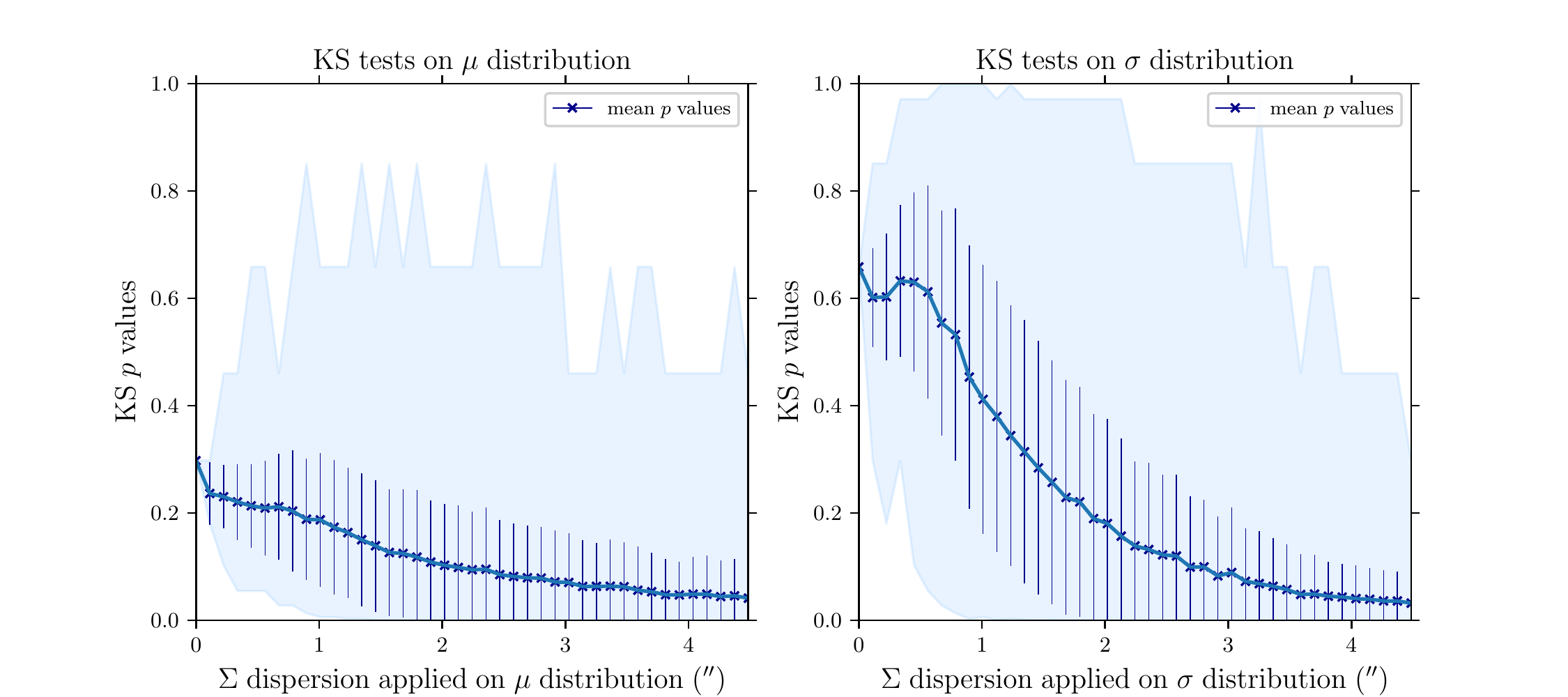}}
\vspace{-0.3cm}
\caption{\small Results from the two-sided 1D and 2D KS tests between the $P$ and C$_2$H moment 0 maps of B335, performed with $N$=100 and $n$=1000. 
\textit{Top row} (left and middle):  $P$ and C$_2$H moment 0 emission maps in color scale, respectively. Both maps are overlaid with the $N$ lines, where the white lines correspond to the $N'$ sample of selected profiles.  The red lines correspond to the profiles where the S/N selection criteria was not reached, and the black lines correspond to the profiles where the 1D Gaussian fit does not converge. (right): Obtained ($\sigma$,$\mu$) values from these 1D Gaussian fits in the form of a 2D diagram, with a linear regression line fitting both the $\sigma$ and $\mu$ values. 
\textit{Middle row:} Results of the two-sided 2D KS tests, where each pixel corresponds to a given Gaussian distribution of width $\Sigma$, added on the sample of $\sigma$ values ($x$-axis) and $\mu$ values ($y$-axis), obtained from the $P$ map. The gray scale corresponds to the $p$-values obtained after $n$ realizations. The left, middle, and right panels, show the mean $p$-value minus the standard deviation of $p$-values, the mean $p$-value, and the mean $p$-value plus the standard deviation of the obtained $p$-values, respectively. The blue contours correspond to $p\,=\,0.2$.
\textit{Bottom row:}   Results of the two-sided 1D KS tests performed on the $\mu$ values (left panel) and on the $\sigma$ values (right panel). In each panel the solid line and markers represent the resulting mean $p$-value as a function of $\Sigma$, which is the width of the Gaussian distribution added on the $\sigma$ and $\mu$ values obtained from the $P$ map. At each point, the shaded area shows the complete range of obtained $p$-values, while the error bars indicate the standard deviation of the distribution of $p$-values after $n$ tries.
Given our $p$ rejection criteria of 0.2, we can infer that the two ($\sigma$,$\mu$) distributions between the C$_2$H moment 0 and $P$ maps could come from a common parent distribution.
}
\label{fig:obs_b335_KS}
\end{figure*}

We present a quantitative comparison between the spatial distribution of the UV-sensitive tracer C$_2$H and that of the polarized dust emission. A simple flux to flux comparison did not appear to be the most relevant method to compare the morphology of the moment 0 of the C$_2$H molecular emission and the polarized intensity $P$. We chose to use two-sided Kolmorogov-Smirnov (KS) tests between these two maps. Only B335 and L1448 IRS2 have high enough sensitivity in both the C$_2$H and $P$ maps for the emissions to overlap sufficiently, the main condition to run the KS tests. Therefore, we performed the tests only on these two sources. In order to match the spatial frequency of emission, both datasets were reduced again using a common uv-range. The C$_2$H moment 0 and $P$ maps were then smoothed to the same angular resolution, and regridded to a Nyquist pattern (4 pixels per beam surface area). 

We now describe the method used to characterize the morphology of the emission in both maps. We follow the propagation path of photons emanating from the central region, and investigate whether a common cause (\ie the irradiation field) can be responsible for both the emission of UV-sensitive molecules and the polarized dust emission. To do so, we draw a number $N$ of lines starting from the peak of the total intensity map and going up to the outer edge of the maps over 360$^\circ$ to obtain a set of 1D emission profiles along those $N$ lines. Among the profiles that have a sufficient number of high signal-to-noise (S/N) pixels in both the C$_2$H moment 0 and polarized intensity maps (we require in this selection that at least five pixels have S/N $\geq$ 3), a 1D Gaussian  fit of the profile is attempted. If the 1D Gaussian fits converge in both the C$_2$H moment 0 and $P$ maps, we retain two parameters, the standard deviation $\sigma$ and the position of the center of the Gaussian $\mu$, for each map. We now have a set of $N'$ (with $N'\leq N$, given that we   selected   profiles for S/N and 1D Gaussian fit convergence reasons) couples ($\sigma$,$\mu$) for both the C$_2$H moment 0 and $P$ maps. 

Before computing the KS tests, we draw sets of perturbations (\ie adding dispersion in the values) from a Gaussian distribution of a given width $\Sigma$ on these ($\sigma_i$,$\mu_i$)$_{i \in\llbracket i,..,N'\rrbracket}$, ($\sigma_i$)$_{i \in\llbracket i,..,N'\rrbracket}$, and ($\mu_i$)$_{i \in\llbracket i,..,N'\rrbracket}$ distributions obtained with the $P$ map\footnote{The same results are obtained if we apply these dispersions on the ($\sigma$,$\mu$) values derived from the C$_2$H map.}. We then perform two-sided 2D KS tests between $\{ (\sigma_i ,\mu_i)_{i \in\llbracket i,..,N'\rrbracket} \}_{\textrm{C}_2 \textrm{H}}$ and $\{ (\sigma_i ,\mu_i)_{i \in\llbracket i,..,N'\rrbracket} + \textrm{T}(\Sigma) \}_{P}$, where $\textrm{T}(\Sigma)$ is a Gaussian distribution of width $\Sigma$. We also perform two-sided 1D KS tests between $\{ (\mu_i)_{i \in\llbracket i,..,N'\rrbracket} \}_{\textrm{C}_2 \textrm{H}}$ and $\{ (\mu_i)_{i \in\llbracket i,..,N'\rrbracket} + \textrm{T}(\Sigma) \}_{P}$, and between $\{ (\sigma_i )_{i \in\llbracket i,..,N'\rrbracket} \}_{\textrm{C}_2 \textrm{H}}$ and $\{ (\sigma_i )_{i \in\llbracket i,..,N'\rrbracket} + \textrm{T}(\Sigma) \}_{P}$. For each value of $\Sigma$ (tested in the range 0-4$''$) we repeat this process $n$ times.

To evaluate if they are drawn
from a common parent distribution, KS tests are performed with these distributions. Each test results in a value $p$, with low values corresponding to different distributions. We reject the hypothesis that two distributions are drawn from the same parent distribution when $p$ < 0.2 \citep{Peacock1983,Fasano1987,Press2007}. At a given $\Sigma$ we evaluate the distributions of the $n$ $p$-values obtained (\ie we derive the mean $p$-value and the standard deviation).

Figure \ref{fig:obs_b335_KS} presents the results of these 1D and 2D KS tests toward B335 (see Appendix \ref{sec:app_KS_tests} for the tests performed on L1448 IRS2), with $N=100$, $n=1000$. The $N'$ S/N selected profiles and corresponding Gaussian fits from the C$_2$H moment 0 and $P$ maps are shown in Figure \ref{fig:obs_b335_1D}. In the  top left and top middle panels of Figure \ref{fig:obs_b335_KS}, the $P$ and C$_2$H moment 0 maps are overlaid with the $N$ profiles respectively, where the white lines represents the $N'$ sample of selected profiles (the red lines correspond to the profiles where the S/N selection criteria was not reached, and black lines correspond to the directions where the 1D Gaussian fit does not converge). The top right panel shows the obtained ($\sigma$,$\mu$) sample in the form of a 2D diagram. The middle and bottom rows show the results of the 2D and 1D KS tests, respectively. Given our $p$ rejection criteria of 0.2, our 2D KS tests performed on B3335 suggest that the two distributions of ($\sigma_i$,$\mu_i$)$_{\textrm{C}_2 \textrm{H}}$ and ($\sigma_i$,$\mu_i$)$_{P}$ may be drawn from the common parent distribution when we consider dispersion values of $\Sigma\,\lesssim1.4^{\prime\prime}$ and $\Sigma\,\lesssim1.8^{\prime\prime}$, applied on the $\sigma$ and $\mu$ distributions, respectively (see the blue contour in the central panel of Figure \ref{fig:obs_b335_KS}). For L1448 IRS2, the same conclusion can be made for $\Sigma\,\lesssim1^{\prime\prime}$ and $\Sigma\,\lesssim1.2^{\prime\prime}$, applied on the $\sigma$ and $\mu$ distributions, respectively (see Figure \ref{fig:obs_L1448_KS}). The selected $N'$ lines, 38 lines in B335 and 43 lines in L1448 IRS2, lie mostly toward the outflow cavity walls of the two protostars, and cover most of the regions where both emission (polarized intensity and C$_2$H moment 0 maps) are sufficiently high. Therefore, we propose that the spatial distribution of the dust polarization and C$_2$H molecular emission maps can have a common origin, assuming that our approach describes  the morphology of these emissions sufficiently well.

In order to investigate the role of the irradiation field in the observed chemistry and the efficiency of the grain alignment, we would need to build models simulating the involved physical and chemical processes. While we perform the modeling of conditions leading to dust grains to align with different efficiencies in Paper II, the modeling of the effect of irradiation on the chemistry is beyond the scope of this work and is left for a future study. We restrict ourselves here to this quantitative comparison, which suggests that the morphology of the $P$ and C$_2$H moment 0 emission maps, are comparable in B335 and L1448-IRS2.

\section{Discussion}
\label{sec:p3_disc}

\subsection{Dust grain polarization and tracers of irradiated molecular gas}
\label{sec:KS_disc}

The spatial distribution of the polarized dust emission and C$_2$H molecular emission suggests that the heating occurs preferentially along the outflow cavities. The accretion activity onto the central protostellar embryo likely represents the major source of photons responsible for  triggering the observed warm carbon chain chemistry \citep{Sakai2009,Sakai2013b} and for  increasing the grain alignment efficiency by extending the population of dust grains subject to alignment by the radiative torques \citep{LeGouellec2020}. These processes can explain the highly polarized dust emission and C$_2$H molecular emission observed toward the dense walls of outflow cavities, as suggested by our statistical tests described in Section \ref{sec:p3_obs_KS}. However, we note that this  analyzis relies on two sources, and other Class 0s presented in Sect. \ref{sec:p3_obs} may show different behaviors.

Several reasons can justify why, in some regions of Class 0s protostellar envelopes, the spatial distribution of the irradiation tracers and polarized dust emission appears to differ.  The morphology of the inner envelope density structures directly dictates the propagation scheme of energetic photons emanating from the central source. For example, Serpens Emb 8 and Serpens Emb 8(N)  exhibit drastic morphological differences in the ALMA dust continuum and polarized dust emission \citep{Hull2017a,LeGouellec2019a}. Serpens Emb 8(N) appears much more axisymmetric than Serpens Emb 8, which can significantly affect the map of the radiation field in the inner core. The disorganized structures of Serpens Emb 8 could thus favor the propagation of the radiation field throughout the entire inner envelope rather than  along outflow cavities, which appear preferentially polarized in Serpens Emb 8(N). In addition, the various timescales of the physical and chemical processes involved here (\ie grain alignment and warm carbon chain chemistry timescales) may vary from one source to another. For example, the precession timescale of the IRAS4A2 outflow \citep{Chuang2021} may prevent the warm carbon chain chemistry from developing farther out in the cavities, explaining why the spatial distribution of the C$_2$H molecular emission is limited compared to the polarized dust emission. In Serpens Emb 8 the increase in irradiation due to the recent accretion burst suggested by \citet{Greene2018} can explain why the C$_2$H molecular emission appears so extended and bright. However, the polarized dust emission within the envelope exhibits a sparse distribution and no significant detection down to the detection levels allowed by the sensitivity of the observations. This could be due to either a more disorganized magnetic field along the line of sight (e.g., \citealt{Valdivia2022}) or to a mean magnetic field orientation closer to the line of sight. Finally, the last element that is crucial while investigating the role of irradiation in the chemistry and grain alignment, is the properties of dust grains. The structure and size distribution of dust grains could vary throughout the envelope, depending on their spatial origin and the radiation field strength, as dust grains can become rotationally disrupted by radiative torques \citep{Hoang2020b}. This  point is further explored in Section \ref{sec:disc_kine} and in our Paper II. For example, this argument could explain why the northern equatorial mid-plane of B335 is so polarized.

\subsection{Contribution of shocks to illumination in cavities}
\label{sec:disc_shocks}

Ultraviolet photons can be produced by mechanical processes, such as at bow shocks in outflows and jets or potentially at smaller shocks along the outflow cavity walls. Herschel observations of far-infrared spectral lines of CO and H$_2$O    revealed evidence of irradiated shocks located along outflow cavity walls, associated with a hot ($\ge\,500K$) gas component \citep{Kristensen2017,Karska2018}. Observations of SO (see Section \ref{sec:p3_obs}), a candidate tracer of the shocks developed at the interaction zones between the outflow--jet system and the ambient envelope gas (\citealt{Lefloch2005}), also suggest the presence of shocks in outflow cavities. These shocks could be an additional source of UV photons \citep{Neufeld1989a,Neufeld1989b} propagating in the cavities \citep{vanKempen2009}. SO can also trace the weak accretion shock of the infalling gas onto the circumstellar disk \citep{Sakai2014a,Sakai2014b,Miura2017,vanGelder2021}. It originates from reactions of atomic S released from the grains with OH, and from H$_2$S converted to SO with atomic oxygen and OH. The fact that we notice spatial overlap of SO emission with dust polarization in the outflow cavities of IRAS4A, Serpens Emb 8, and Serpens 8(N) (see Section \ref{sec:p3_obs}) tends to suggest that the shock activity plays a role in the alignment of dust grains (see, e.g., \citealt{Hoang2019d}). To quantify the amount of self-irradiation in the candidate shocks, one possibility could be to model the surface geometry of the cavity walls (see a similar work performed for a bow shock in \citealt{Gustafsson2010}),  and to implement self-consistently the shock models that account for self-irradiation \citep{Lehmann2020}.

\subsection{Conditions   for  development of UV-irradiated gas and efficient grain alignment in protostellar envelopes}
\label{sec:disc_UV}

\citet{LeGouellec2020} showed that the grain alignment efficiency is high and roughly constant with respect to column density in a prototypical Class 0 protostellar core. However, they could not entirely reproduce it; their radiative transfer calculations show an alignment efficiency that is too low. However,   irradiation seems to be an important factor that drives the extent of the regions where grain alignment  produces polarized dust emission efficiently (\ie where the irradiation is high enough to align grains). The wide variety of the observed spatial extent of the molecular emission lines we presented in Section \ref{sec:p3_obs} can bring additional clues   to the nature and morphology of the radiation field in the inner core.

\subsubsection{Using UV-sensitive chemistry and its associated radiation field to explain the observed ALMA dust polarization maps}

Interactions between UV radiation from shocks in jets and/or outflows and in cavity walls (see above) and the ambient material could cause the growth of regions of PDR-like conditions \citep{LeeS2015}. 
The identification of PDR tracers toward outflow cavity walls suggests that the UV radiation field is significant in these regions. As mentioned above, one of the limitations of our radiative transfer calculations is the fact that the radiative energy derived in each cell only originates from the reprocessing of the radiative energy escaping from the sink. Analyzing the UV-sensitive chemistry of the inner core could allow us to confirm or rule out the radiation field we implement in our models.

The molecule C$_2$H is usually seen in  PDRs, for example  the Orion Bar \citep{Teyssier2004,Pety2005,vanderWiel2009,Nagy2015} and the Horsehead Nebula \citep{Cuadrado2015,Guzman2014,Guzman2015}, located at the irradiated, and thus warmer, edge of these regions. C$_2$H was also found enhanced in the presence of UV radiation toward molecular clouds \citep{Fuente1993,Hogerheijde1995,Jansen1995}. This molecule involves C, C$^+$, and CH$^+$ in its chemical formation pathway, which is maintained in high abundance in the gas phase in the presence of strong UV radiation \citep{Stauber2004,Benz2016}. C$_2$H could thus be expected to trace the (UV-irradiated) outflow cavity walls. C$_2$H has been detected in a variety of Class 0/I protostellar cores, especially toward outflow cavity walls \citep{Jorgensen2013,Oya2014,Imai2016,Higuchi2018,Oya2018,Okoda2018,Bjerkeli2016c,ZhangY2018,Bergner2020,LeeS2020,Okoda2020,Tychoniec2021,Ohashi2022}. In particular, \citet{Murillo2018} exhibited the anti-correlation between the cold gas tracer DCO$^+$ \citep{Murillo2015,LeGouellec2019a} and the warm irradiated regions traced by C$_2$H (and also $c$-C$_3$H$_2$) in two Class 0 cores.
Finally, due to an increase in atomic C and C$^+$ in the gas phase, C$_2$H is also commonly observed and modeled at the upper layer of protoplanetary disks where UV and X-ray photons from the stellar irradiation field can propagate more easily \citep{Henning2010,Kastner2014,Kastner2015,Bergin2016,Cleeves2018,Kastner2018,Bergner2019,Miotello2019,Cleeves2021}. Another candidate mechanism to explain the presence of carbon chain molecules in the gas phase is the top-down destruction of polycyclic aromatic hydrocarbons (PAHs) photo-eroded by the radiation field \citep{Teyssier2004,Pety2005,vanderWiel2009,Pety2012,Guzman2015}. However, this proposition is  challenged by the lack of observational detections of PAHs in protostellar cores \citep{Geers2009}.

The presence of C$_2$H (but also $c$-C$_3$H$_2$ and CS; see Appendix \ref{sec:app_other_lines}) toward the outflow cavities and cavity walls thus suggests that a significant UV radiation field can survive in the inner $\sim$500 au region of Class 0 protostellar cores. 
Given the average penetration path length of UV photons through the dense cavity walls, these photons are reprocessed quickly. 
These reprocessed photons can still contribute to the alignment of small grains, as the mean wavelength of the corresponding radiation spectrum is shorter than the longer-wavelength photons responsible for the alignment of dust grains with sizes $\ge\,1\,\mu$m. 
The equatorial mid-planes, which   presumably do not experience a similar radiation field (the equatorial mid-planes B335 are depleted in C$_2$H emission line for example), would be dense enough to be protected by rotational disruption (depending of the opacity of the inner circumstellar disk; see Paper II), but would require large grains to produce the observed polarized dust emission. 
The case of Serpens Emb 8 is, however, paradoxical. The C$_2$H emission that peaks all over the inner core could be due to significant leaks of irradiation from the very disorganized envelope. In this source dust grains might be rotationally disrupted along a variety of directions, which, along with the disorganized magnetic field orientations, can explain the undetected dust polarization in a large portion of the dense regions.

Previous chemical studies   constrained the irradiated field in Class 0 protostellar cores. On scales of $\sim\,1000$ au, comparing atomic line fluxes such as [CII] or [OI] with PDR models, \citet{Karska2018} found that for typical low-mass protostars the UV field was $\sim\,$10$^{2}$ G$_0$ (where G$_0$ is the  radiative flux in the range $\sim$ 90-200 nm), and 10$^{3}$ G$_0$ for protostars in the regime of  SMM1, the brightest protostar in Serpens Main with $L_{\textrm{bol}}\,\gtrsim\,120$ \citep{Enoch2011,Kristensen2012}. Using UV-irradiated shocks models, they found a UV field of 1-10 G$_0$ using pre-shock densities of 10$^{5}$ cm$^{-3}$. In addition, comparing the observed fluxes and line ratios of several ionized hydrides such as C$^{+}$, CH$^{+}$, OH$^{+}$, H$_2$O$^{+}$, and HCO$^{+}$ with chemical models, \citet{Benz2016} constrained values of 200-400 G$_0$ for NGC1333 IRAS4A and 2-8 G$_0$ for SMM1 at the \textit{Herschel} half-power beam radius at 300 $\mu$m, which corresponds to a radius of 3000 and 5000 au from the center of the protostar at the distance of Perseus and Serpens, respectively. 
Using the  IRAM 30m observations, similar G$_0$ values were derived in Serpens Main by \citet{Mirocha2021}, who computed ratios of CN to HCN column densities, a good tracer of UV fields around around low- and intermediate-mass protostellar cores.
In our radiative transfer calculations, the radiation field spectrum computed in these regions has been entirely reprocessed toward longer wavelength photons (IR to millimeter wavelengths), which means the G$_0$ is negligible there. 

As suggested above, in outflow cavities the high-energy (UV) part of the radiation field spectrum must originate from the shocks occurring between the outflowing gas and the envelope. Therefore, it is not possible to precisely compare the radiation field obtained from radiative transfer calculations (where the radiation field emanating from the center is heavily reprocessed) with these chemical studies that only discuss the impact of the UV field on the chemistry. For reference, at 1000 au from the protostar our fiducial model  (see Paper II) yields  an irradiation field of $u_{\textrm{rad}}/u_{\textrm{ISRF}}\,\simeq\,3\times10^{3}$ in the direction of the outflow and $\sim\,5\times10^{2}$ in the direction of the equatorial planes. At 3000 au from the protostar the ratio $u_{\textrm{rad}}/u_{\textrm{ISRF}}$ is $\sim\,3.5\times10^{2}$ in the direction of the outflow and $\sim\,1\times10^{2}$ in the direction of the equatorial planes. The resulting grain alignment efficiency does not seem to be high enough to reproduce the results obtained with ALMA observations obtained in \citet{LeGouellec2020}. 
In addition, the higher values of irradiation we explore may favor the rotational disruption of grains and thus deteriorate the conditions for grain alignment removing the large grains. This latter point is quantitatively addressed in our Paper II. 
While irradiation appears to locally affect the grain alignment conditions, the global efficiency of the grain alignment mechanisms is also important to take into account. \citet{LeGouellec2020} also noted that the paramagneticity   of dust grains strongly affects the grain alignment efficiency. 
Recently detailed modeling by \citet{Hoang2022}, \citet{Hoang2022b}, and \citet{ChauGiang2022} have shown that to ensure the alignment of large dust grains ($\ge$10\,$\mu$m; necessary to reproduce the observed polarized dust emission), grains containing embedded iron inclusions are required to ensure efficient grain internal and external alignment processes with respect to the gas randomization (\ie short Barnett relaxation and Larmor precession timescales with respect to the gaseous damping timescale).




\subsubsection{Effect of  C/O  in the gas phase and the CCH abundance}
In protoplanetary disks only models implementing C/O>1 can reproduce the high abundance of C$_2$H \citep{Bergin2016,Kama2016,Cleeves2018,Kastner2018,Miotello2019,Alarcon2020,Anderson2021,Cleeves2021,Facchini2021}, otherwise CO would carry most of the carbon content. In disks, to obtain a high C/O ratio, one possible explanation is the enhancement of carbon via destruction (and/or photodesorption) of carbon grains and PAHs \citep{Anderson2017}, while a deficit in oxygen can be caused by grain growth and dust settling \citep{Hogerheijde2011,Du2015,Salinas2016,Du2017,Cleeves2018}. However, \citet{Bergner2020} did not find that C$_2$H formation is favored by CO depletion in a few of Class 0/I systems, suggesting that CO can   already be quite depleted in the gas phase. 
In outflow cavity walls, PDR-like modeling is required to fully understand the exact role of the UV radiation field in the enhancement of C$_2$H, but the C/O ratio is a parameter that may also play a role in the C$_2$H enhancement, in this active region separating the warm outflow material from the dense cold infalling envelope. In this region the temperature gradient is expected to be large because of the large gas density gradient across cavity walls and reprocessing of the local irradiation field. It is possible that in these specific conditions, the gas experiences a rather sharp transition from cold envelope conditions to temperatures allowing the CO sublimation out of dust grains \citep{Anderl2016}. C$_2$H might be detected where a high C/O ratio could be explained by a limited CO sublimation. This is illustrated by the dichotomy found in BHR71 IRS2 between the outflow cavity and inner envelope traced by CO and C$^{18}$O, respectively, and the cold envelope traced by N$_2$D$^{+}$ where CO freezes out onto dust grains (\citealt{Hull2020a}; see also the unpolarized southern region of B335 in Figure \ref{fig:obs_b335_chem} compared to these same molecular lines presented in \citealt{Cabedo2021,Cabedo2022}). The dust polarization detected in the outflow cavity walls of IRS2 lies precisely  between the regions where the  N$_2$D$^{+}$ and CO emission lines peak, suggesting that dust grains cannot become aligned in these areas. This can be explained by the attenuation of the irradiation field, which is able to align far fewer grains. In addition, another cause of the drop in polarized intensity where CO freezes out onto grains could be that CO ices on grain surfaces decrease the helicity of the  dust grains, and thus the polarization. Modeling dust grain surfaces and volatile chemistry toward the outflow cavity walls will be required to constrain the role played by the  C/O ratio and the radiation field spectrum and strength in the observed warm carbon chemistry, and potentially the grain alignment conditions.


%



\subsubsection{Understanding the origin of  polarized dust in cavity walls via gas kinematic studies}
\label{sec:disc_kine}

The regions traced by C$_2$H and CS around the lobes of bipolar outflows are thought to correspond to outflow cavities walls, because of their irradiated nature. Assuming that   the polarized dust emission and the UV-sensitive molecular emission lines both come from these cavity walls, it is necessary  to characterize the kinematic nature of outflow cavity walls to further constrain the physical conditions that favor grain alignment. \citet{ZhangY2016} have proposed  tentative evidence of a slowly moving rotating outflow using CS, suggesting that cavity walls are entrained by a rotating disk wind launched at large radii from the source. Further evidence of rotation toward Class 0 objects was found by \citet{ZhangY2018} and \citet{Ohashi2022} using CS and C$_2$H. The authors argue that these molecules should trace the gas entrained by the outflow along the cavity walls, or the magneto-centrifugal wind itself launched from the disk, rather than the gas infalling from the envelope to the central object. Even if the rotation of the outflow cavity walls is minimal, it has been suggested from observations of a few objects that the specific  angular momentum of the corresponding gas is greater than that of the infalling-rotating envelope \citep{Oya2018,ZhangY2018}. We note, however, that \citet{Okoda2018} have concluded that the C$_2$H emission in a warm carbon-chain chemistry (WCCC) protostar was preferentially tracing  gas from the infalling-rotating envelope. They attribute the C$_2$H emission, also detected around outflow cavity, to the infalling-rotating envelope. 

In Class 0 protostars it is thus still unclear whether the emission from the C$_2$H molecule originates from the regions surrounding outflow cavities rather than from the outflowing gas itself. However, outflow cavity walls regions are  subject to a significant shear. The material located at the interface between the outflowing gas and the infalling envelope material must experience turbulent mixing. The mixing layer is replenished by outflowing material and infalling envelope at a rate that is proportional to the mass outflow rate, but the mass brought to the mixing layer by the outward shocked wind should be larger than the mass brought by the infalling envelope (see   \citealt{Canto1991,Liang2020}, where this result was quantified via analytical models). \citet{Liang2020} developed a model of a steady-state cavity caused by the interaction of a wind with an infalling rotating envelope, whose size is dictated by the equilibrium between the wind ram pressure and envelope thermal pressure. They were able to match their predictions of the warm turbulent mixing layer with the broad-line profile of high-$J$ CO line observed with Herschel toward Serpens SMM1 \citep{Kristensen2017b}. In this source the mixing layer is made of warm gas (T$\,\sim\,$250 K), due to the competition between turbulent heating, expansion, and CO cooling. 

It is still unclear whether the regions of high dust polarization and C$_2$H emission belong to the outflowing gas, the mixing layer, or the infalling envelope. This question is of great importance given the local physical conditions required by grain alignment, especially the maximum size of aligned dust grains \citep{Valdivia2019,LeGouellec2019a,Hull2020a}. To discuss the presence of large grains ($\ge$10\,$\mu$m) in cavity walls,  we must address the question of where and when the required dust grain growth has occurred. Dust growth may be efficient in situ, within the infalling material of the protostellar envelope (\citealt{Galametz2019,Bate2022}; infalling structures such as streamers may also favor dust grain growth thanks to their higher density). However, if the material producing the polarized dust emission in outflow cavity walls is entrained by a disk wind, large dust grains in cavities could originate from the much denser regions of the inner circumstellar disk where the outflow originates, and where grains grow more quickly. Recent studies have investigated the possibility of dust ejection from the inner disk \citep{Wong2016,Liffman2020,Vinkovic2021,Hutchison2021,Booth2021,Tsukamoto2021,Rodenkirch2022}. Kinematic studies via observations of the spectral lines tracing those highly polarized regions could thus clarify which of these two scenarios can explain the spatial origin of these large grains.

Additionally, mechanical grain alignment may also play a role in regions harboring anisotropic (supersonic) gas-dust drift velocity streams. Originally, the Gold alignment \citep{Gold1952} was proposed to align the grains with long axes along the flow in supersonic conditions. More recently, the alignment of irregular grains induced by gaseous collisions was introduced in \citet{Lazarian2007}. This mechanical torque (MET) alignment mechanism acts similarly to that caused by  RATs, where mechanical torques are applied by gas particle momentum instead of the photon momentum.
Recent numerical quantification have shown a large range of grain spin-up efficiencies by METs for different grain shapes and gas-dust drift velocities \citep{Hoang2018,Reissl2022}.
Alignment would occur with respect to the orientation of the average gas-dust drift velocity component or the magnetic field, depending on the amount of the iron inclusions and the gas-dust drift velocity \citep{Lazarian2021,Hoang2022b}.
In protostellar environments, unlike the infalling streams of gas located within the envelope, the outflow cavity walls are   next to high velocity outflowing gas. The shocked layers defining the interaction between the outflowing gas and the envelope may exhibit favorable conditions for the mechanical alignment of dust grains. However, the width of the shock fronts would   likely be  small compared to the angular resolution of the observations presented here (see, \eg  \citealt{Mottram2014,Kristensen2017b}), such that the polarized dust emission would not be dominated by the shock front. The turbulent mixing layer (see \citealt{Canto1991,Liang2020}), marking the transition between the outflowing and infalling velocities, could extend to a larger area. The gas-dust drift would be less anisotropic, but the (sub)millimeter polarized dust emission more likely emanates from this region. 
It is still unclear what   the relative contributions of RATs and METs are in outflow cavity walls.
Revealing this question would require   precisely constraining the gas-dust drift velocities and irradiation conditions undergone by the dust grains responsible for the observed (sub)millimeter polarized dust emission.

\section{Conclusions and summary}
\label{sec:p3_ccl}

We compared the spatial distribution of polarized dust emission with molecular emission line of UV-sensitive species in the envelope of Class 0 protostars (B335, L1448 IRS2, NGC1333 IRAS4A, Serpens Emb 8, and 8(N)). The goal was to use the warm chemistry processes to characterize the role of the radiation field in grain alignment within the dense envelope of protostars. The main results and conclusions of this paper are as follows:

\begin{enumerate}
\item We qualitatively compare the observed maps of dust polarization with maps of the irradiated regions probed by UV-driven chemistry traced by the emission of the C$_2$H hydrocarbon. The emission of this molecule is seen to be enhanced along outflow cavities walls, where in the case of the axisymmetric cores the polarized dust emission is also enhanced.
\item In the case of a more turbulent core, such as Serpens Emb 8, the energetic photons probed by the C$_2$H emission seem to propagate throughout the entire inner core. In this source, the polarized intensity exhibits very localized detections, with only tentative correlations with the location of C$_2$H molecular gas emission. 
\item We show that the shock tracer SO also peaks toward outflow cavities where polarized dust emission is enhanced. Shocks happening along cavity walls can represent an additional source of high-energy photons that may contribute to the anisotropic radiation field aligning grains via radiative torques.
\item In the two protostars presenting the best detection of both the polarized dust emission and the C$_2$H emission, we find a quantitative correlation between the morphology of the polarized intensity and C$_2$H moment 0 integrated emission maps, suggesting that the radiation field impinging on the cavity walls may favor both the grain alignment and the warm carbon chain chemistry in these regions. 
\item However, some regions of Class 0 protostellar envelopes, which are thought to be under a weaker radiation field, still exhibit enhanced polarized intensity, such as the the northern equatorial mid-plane of B335, contrary to the southern one. This may suggest that a fraction of dust grains contributing to the millimeter dust continuum emission may be aligned even in the absence of an ultraviolet radiation field, but with a strongly reddened radiation field.

\end{enumerate}

Comparing the ambient chemistry that develops where dust polarization is observed opens a promising avenue to further constrain grain alignment, given that the modeling of the results from these two fields can constrain  crucial parameters together,  such as radiation field, dust temperature, and the subsequent sublimation radii beyond which gas species can freeze onto dust grains,  potentially modifying their alignment properties. We highlight the importance of source geometry and evolution in setting the environmental conditions of the envelope of protostars, which in turn govern the capability of radiative torques to align dust grains depending on the spectrum of local photons.



~\\
{
\small
\textit{Acknowledgments}
V.J.M.L.G. thanks \L. Tychoniec, N. Sakai, and J. M. Girart for their help and support regarding the ALMA molecular lines and dust polarization observations presented in this paper.
V.J.M.L.G. and C.L.H.H. acknowledge the ESO Studentship Program, and the guidance and support of Eric Villard.
C.L.H.H. acknowledges the support of both the NAOJ Fellowship as well as JSPS KAKENHI grant 18K13586. 
This work has received support from the European Research Council (ERC Starting Grant MagneticYSOs with grant agreement no. 679937).
This paper makes use of the following ALMA data: ADS/JAO.ALMA\#2013.1.01102.S, 
ADS/JAO.ALMA\#2016.1.01552.S, 
ADS/JAO.ALMA\#2017.1.01392.S, ADS/JAO.ALMA\#2013.1.01380.S, ADS/JAO.ALMA\#2018.1.01873.S, 
ADS/JAO.ALMA\#2016.1.01501.S, 
ADS/JAO.ALMA\#2016.1.00604.S, 
ADS/JAO.ALMA\#2013.1.00031.S, 
ADS/JAO.ALMA\#2016.1.01089.S, 
ADS/JAO.ALMA\#2013.1.01102.S, 
ADS/JAO.ALMA\#2016.1.01089.S, ADS/JAO.ALMA\#2015.1.00546.S., 
ADS/JAO.ALMA\#2017.1.01174.S, 
ADS/JAO.ALMA\#2016.1.00710.S, ADS/JAO.ALMA\#2013.1.00726.S, ADS/JAO.ALMA\#2015.1.00768.S. 
ALMA is a partnership of ESO (representing its member states), NSF (USA) and NINS (Japan), together with NRC (Canada), MOST and ASIAA (Taiwan), and KASI (Republic of Korea), in cooperation with the Republic of Chile. The Joint ALMA Observatory is operated by ESO, AUI/NRAO and NAOJ.
The National Radio Astronomy Observatory is a facility of the National Science Foundation operated under cooperative agreement by Associated Universities, Inc.
\\

\textit{Facilities:} ALMA

\textit{Software:} APLpy, an open-source plotting package for Python hosted at \url{http://aplpy.github.com} \citep{Robitaille2012}. CASA \citep{McMullin2007}. Astropy \citep{Astropy2018}. POLARIS \citep{Reissl2016}

}

\bibliography{ms}
\bibliographystyle{apj}


\begin{appendix}
\addcontentsline{toc}{section}{Appendix}
\renewcommand{\thesection}{\Alph{section}}

\clearpage
\section{\normalfont{Additional tracers of warm chemistry in outflow cavities.}}
\label{sec:app_other_lines}

We present here maps of two additional tracers of warm chemistry, CS and  $c$-C$_3$H$_2$ (see Table \ref{t.obs_lines} for the transition information). The molecule $c$-C$_3$H$_2$ has a similar chemistry to that of C$_2$H. Like C$_2$H, it has been detected toward the  PDR edges \citep{Teyssier2004,Pety2005,Cuadrado2015,Guzman2014,Guzman2015}, and is likely an irradiation tracer. 
Detections of $c$-C$_3$H$_2$ have been found at the scale of cores ($\sim\,2000-4000$ au) toward proto-stellar and pre-stellar cores irradiated by external sources, suggesting irradiation driven chemistry (a nearby star, for example, or the interstellar radiation field itself; see \citealt{Lindberg2012,Spezzano2016}).
$c$-C$_3$H$_2$ (and C$_2$H) can also probe the disk--envelope interface, at the limit between the centrifugal barrier and the infalling-rotating envelope \citep{Sakai2014b,Sakai2014a}.
CS is also a molecule whose emission can be enhanced at the edges of PDR \citep{RiviereMarichalar2019}. Commonly seen toward outflow cavities tracing the warm molecular layer \citep{Kwon2015,Bjerkeli2016b,Oya2018,ZhangY2016,MartinDomenech2019,ZhangY2018,Okoda2020,Taquet2020}, CS may trace the UV-irradiated regions of disks as well \citep{Podio2020a,Podio2020b}. We note, however,  in disks seen edge-on that CS can appear below CN, whose production is favored by the irradiation of the disk surface \citep{RuizRodriguez2021}. The chemical formation route of CS may require strong UV radiation as it can trigger reactions between S/S$^+$ and small hydrocarbons to produce carbonated S-species, or photodesorption from dust grains \citep{LeGal2019b,Podio2020b}. In the protostars we study, CS is less obviously tracing the PDR-like outflow cavity wall layer because its spectral emission has a narrower profile than SO, but a broader extent compared to that of C$_2$H that only peaks at the $v_{\textrm{lsr}}$ (see the kinematic study present in Figure \ref{fig:obs_iras4a2_spectra}; see also \citealt{Tychoniec2021,Ohashi2022}). In addition, applying the method presented in Section \ref{sec:p3_obs_KS} on the $P$ and CS moment 0 maps of B335 and L1448 IRS2 does not result in any probing results (especially for L1448 IRS2), suggesting that the morphology of the CS moment 0 map is less linked to outflow cavity walls.

\begin{figure*}[!tbh]
\centering
\hspace{-0.5cm}
\subfigure{\includegraphics[scale=0.618,clip,trim= 0cm 2cm 7.3cm 3cm]{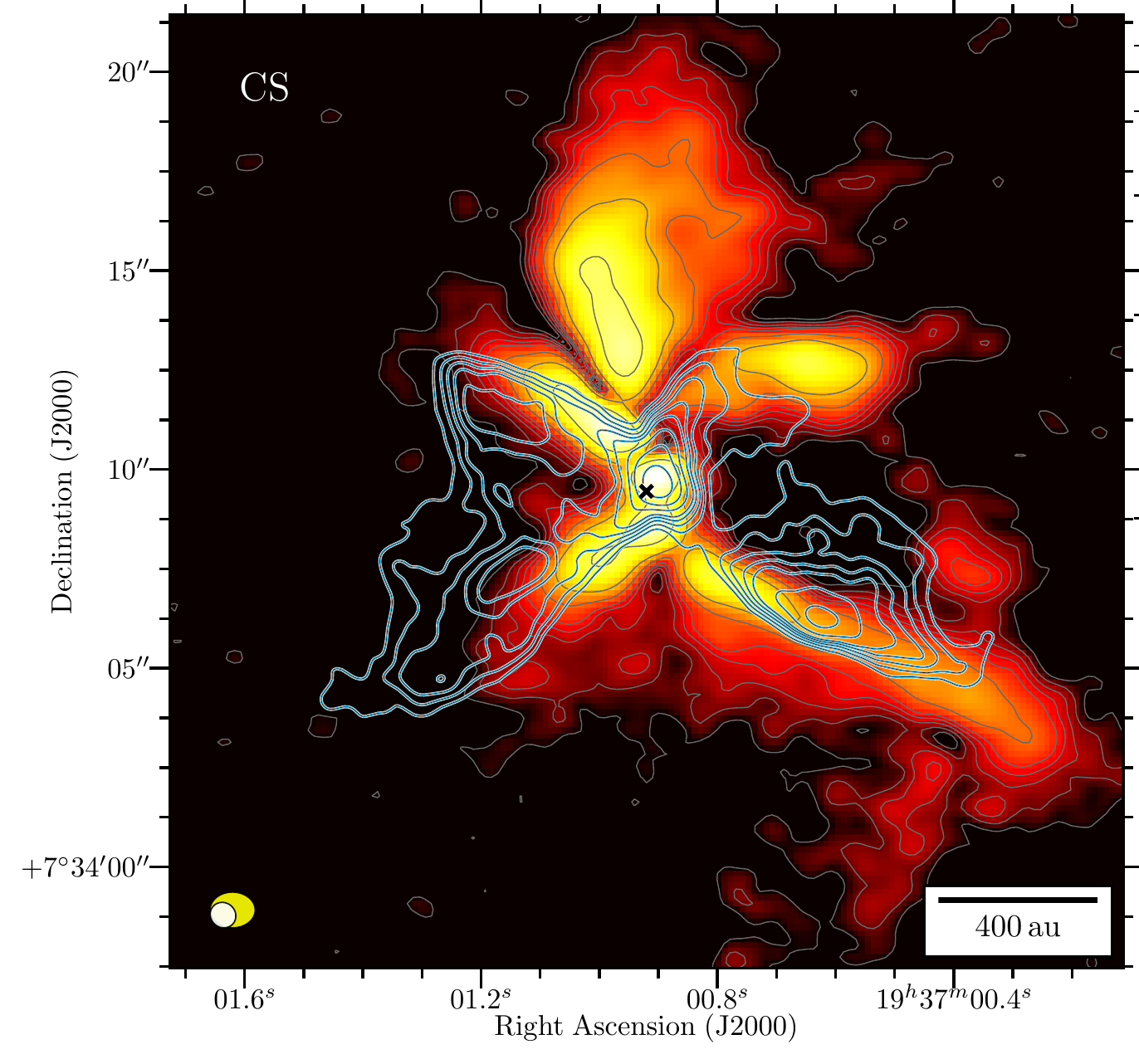}}
\subfigure{\includegraphics[scale=0.618,clip,trim= 2cm 2cm 4cm 3cm]{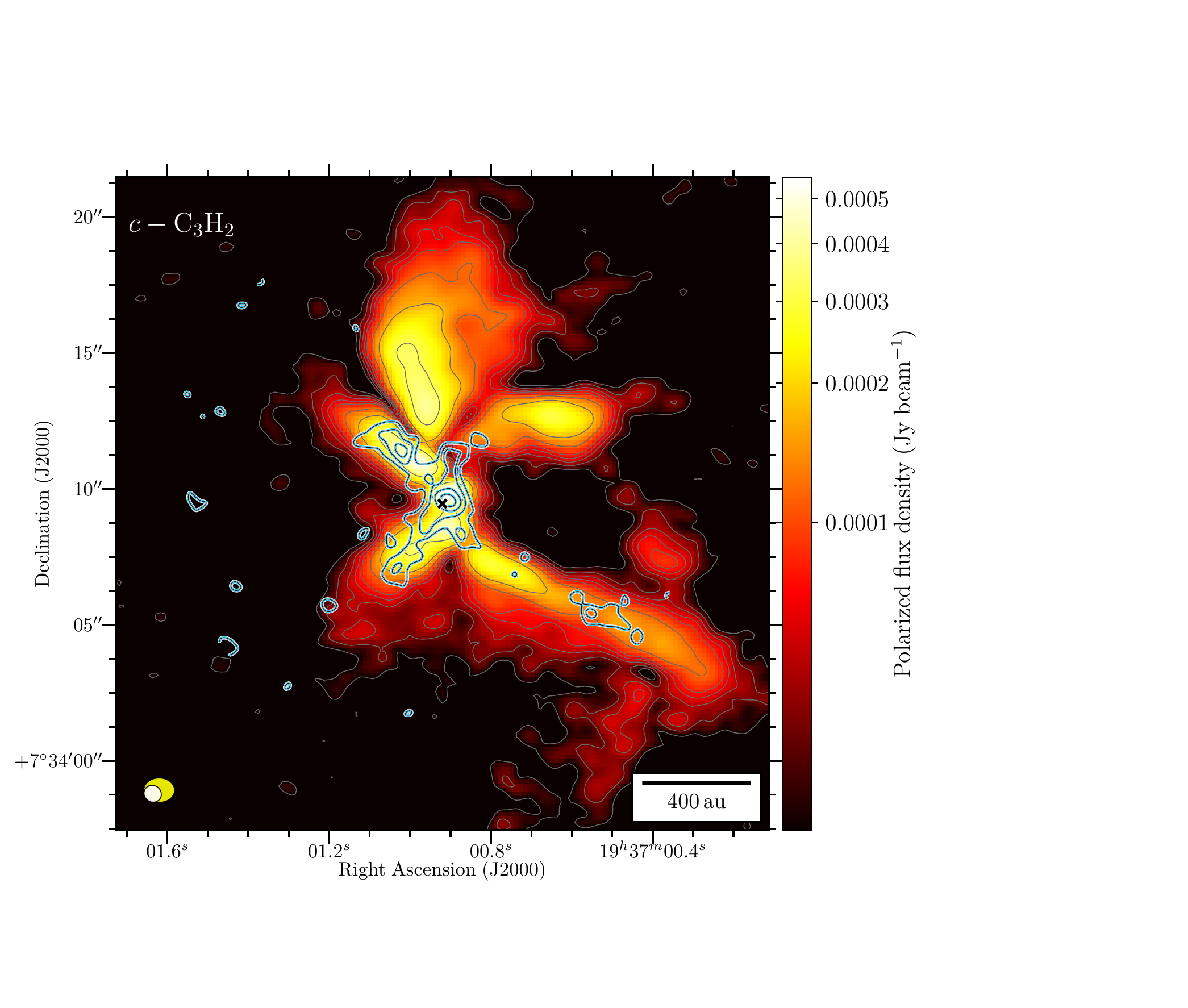}}
\vspace{-0.2cm}
\captionof{figure}{\small Polarized intensity and molecular line observations around the B335 Class 0 protostellar core (as in Figure \ref{fig:obs_b335_chem}). The thick light blue contours trace the moment 0 map of the CS (from 6 to 11 \kms) and $c$-C$_3$H$_2$ (from 7 to 9.5 \kms) molecular emission spectral line, at levels of 3, 5, 7, 9, 11, 16, 24, 44 $\times$ the rms noise level of each map. In the bottom left corner of each plot the  ellipses represent the beam resolution element of a given dataset: yellow for polarized dust emission;  white for the CS and $c$-C$_3$H$_2$ spectral emission lines. ALMA 1.3mm dust polarized emission from \citet{Maury2018} and Maury et al. (in preparation). The CS and $c$-C$_3$H$_2$ emission lines are from \citet{Imai2016}.
}
 \label{fig:obs_b335_chem_app}
\end{figure*}

\textbf{B335:}  Figure \ref{fig:obs_b335_chem_app} presents the $c$-C$_3$H$_2$ and CS moment 0 integrated emission maps from \citet{Imai2016}. The emission of the carbon chain molecule $c$-C$_3$H$_2$, and the CS molecular emission, are clearly enhanced toward the outflow cavities. The $c$-C$_3$H$_2$ emission is very faint, however, and is located exclusively at the base of outflow cavity walls.

\begin{figure}[!tbh]
\centering
\hspace{-0.7cm}
\includegraphics[scale=0.52,clip,trim= 0cm 2cm 4cm 3cm]{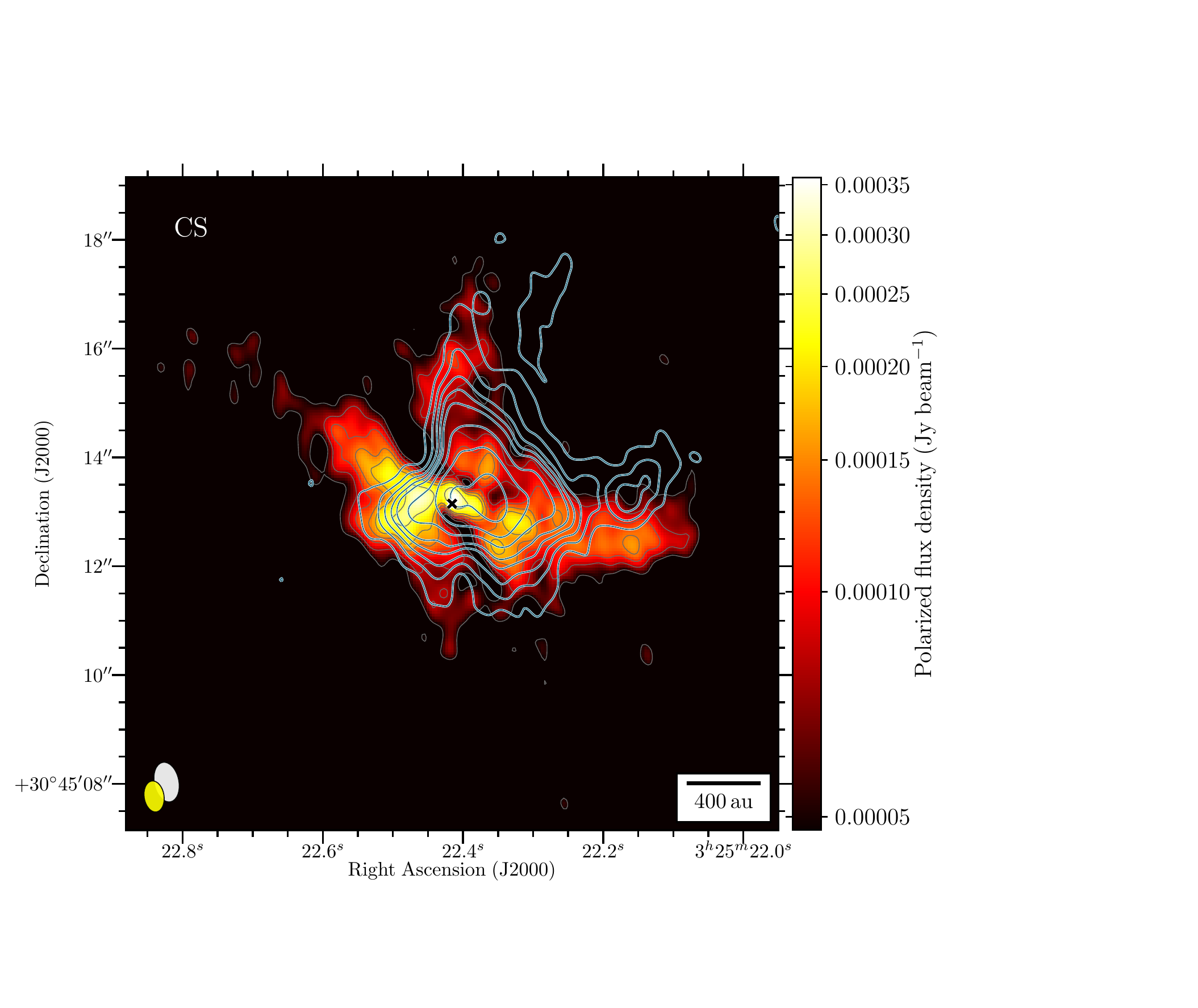}
\captionof{figure}{\small Magnetic fields, polarized intensity, and molecular line observations around the L1448 IRS2 Class 0 protostellar core (as in  Figure \ref{fig:obs_l1448_chem}). The thick light blue contours (top right and bottom panels) trace the moment 0 map of the CS (by integrating emission from 2 to 6 \kms) molecular emission spectral line. The ALMA 1.3mm dust polarized emission is from \citet{Kwon2019} and the CS emission line is from ALMA project 2016.1.01501.S, Y. Zhang (in preparation).
}
 \label{fig:obs_l1448_chem_app}
\end{figure}

\textbf{L1448 IRS2:}  
Figure \ref{fig:obs_l1448_chem_app} presents the CS moment 0 integrated emission map (from Y. Zhang in preparation, ALMA project 2016.1.01501.S, PI: N. Sakai)
The C$_2$H molecular line shown in Figure \ref{fig:obs_l1448_chem}, is more extended than CS, but emission from both molecular lines appears enhanced toward the walls of the cavities.

\begin{figure}[!tbh]
\centering
\hspace{-0.64cm}
\subfigure{
\includegraphics[scale=0.52,clip,trim= 0cm 2cm 4cm 3cm]{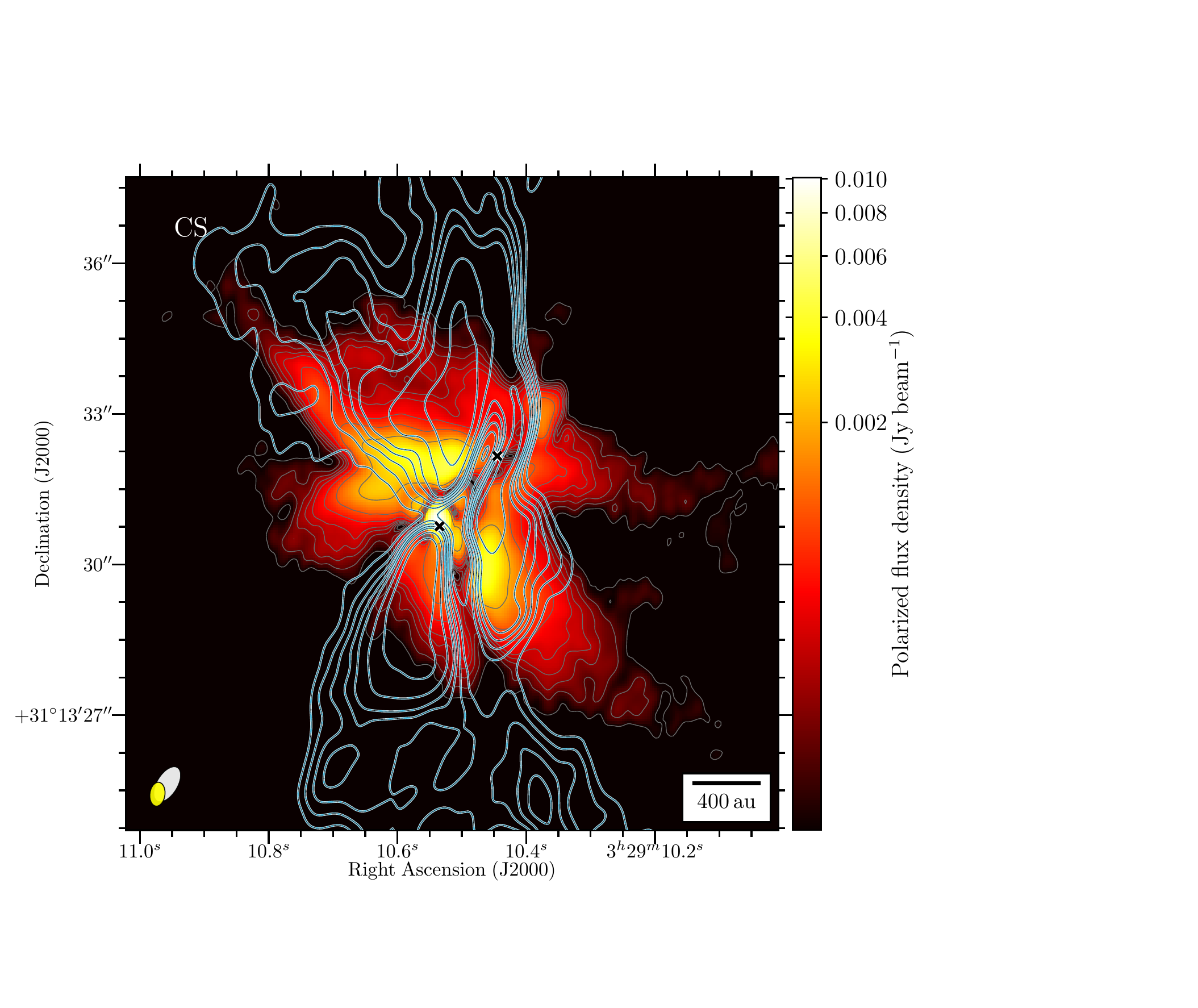}}
\caption[Magnetic fields, polarized intensity, and molecular line observations around the NC1333 IRAS4A Class 0 protostellar core]{\small Magnetic fields, polarized intensity, and molecular line observations around the NC1333 IRAS4A Class 0 protostellar core (as in  Figure \ref{fig:obs_iras4a_chem}). The thick light blue contours (top right and bottom panels) trace the moment 0 map of the CS (by integrating emission from -10 to 19.5 \kms) molecular emission spectral line. The two crosses are located at the two peaks of the dust continuum map. IRAS4A1 is to the southeast of IRAS4A2. The ALMA 1.3mm dust polarized emission is from \citet{Ko2019}. The CS emission line is from \citet{Chuang2021}.
}
 \label{fig:obs_iras4a_chem_app}
\end{figure}

\textbf{NGC1333 IRAS4A:} 
Figure \ref{fig:obs_iras4a_chem_app} presents the CS moment 0 integrated emission map from \citet{Chuang2021}. The polarized dust emission along the outflow cavity walls seems to overlap precisely with the CS molecular emission line in the southern cavity of IRAS4A1 and IRAS4A2, and in the northern cavity of IRAS4A1.

\clearpage
\section{\normalfont{Details about our geometrical comparisons between the maps of polarized intensity and integrated molecular emission line}}
\label{sec:app_KS_tests}

Figure \ref{fig:obs_b335_1D} presents the emission profiles of the S/N selected lines obtained in the C$_2$H moment 0 and $P$ maps of B335 from the method presented in Section \ref{sec:p3_obs_KS}. These profiles correspond to the black and white lines in Figure \ref{fig:obs_b335_KS}.

\begin{figure*}[!tbh]
\centering
\subfigure{
\includegraphics[scale=0.6,clip,trim= 2.5cm 1cm 0cm 0cm]{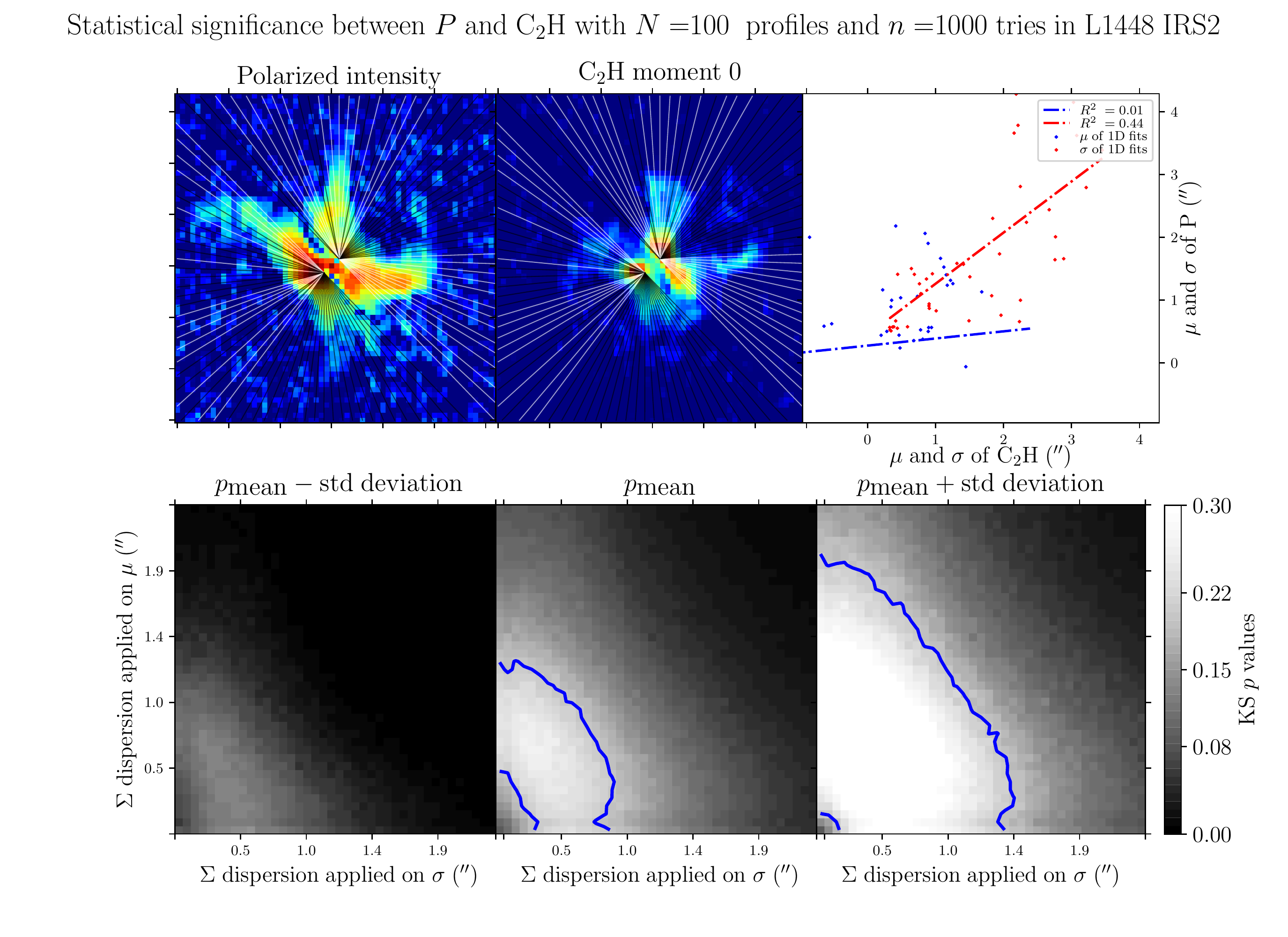}}
\subfigure{
\includegraphics[scale=0.55,clip,trim= 1.5cm 0cm 1.5cm 0.4cm]{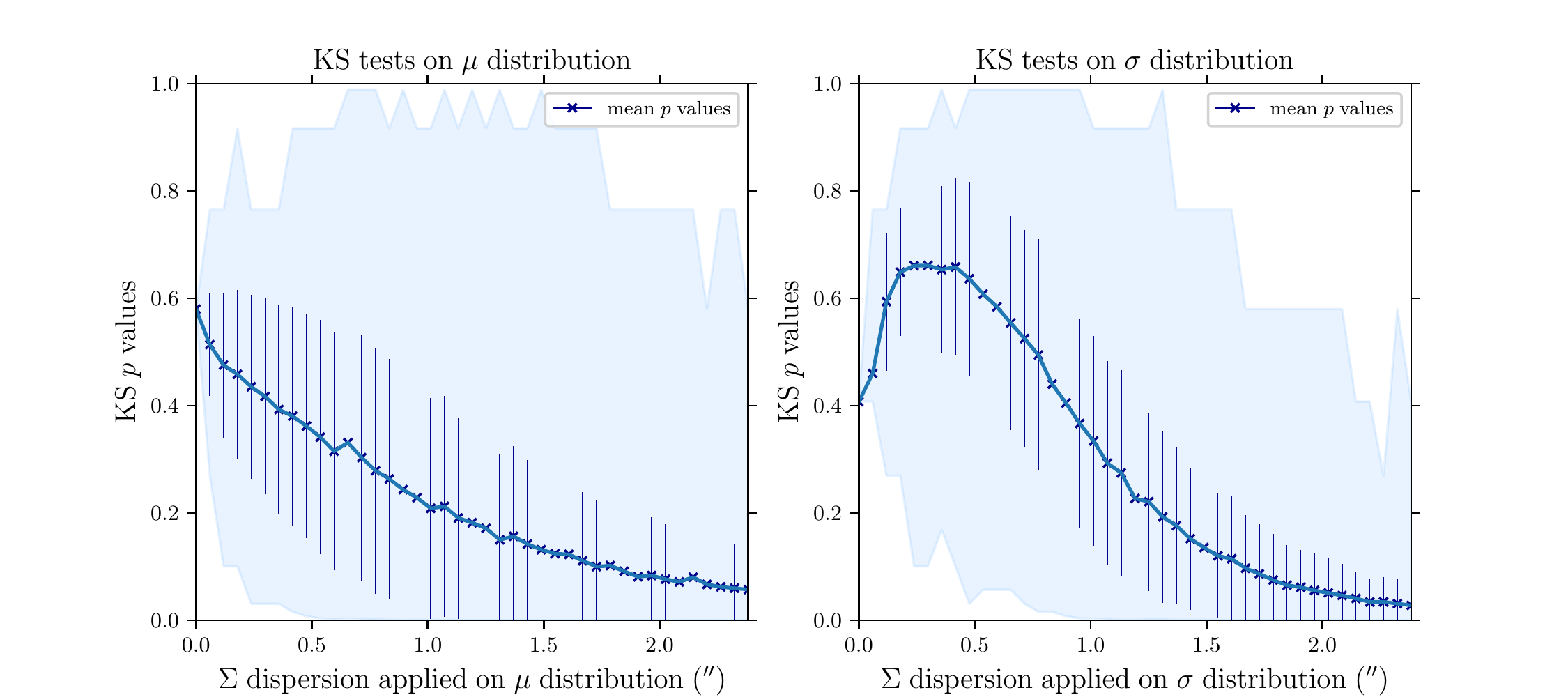}}
\caption{\small Results from the two-sided 1D and 2D KS tests between the $P$ and C$_2$H moment 0 maps of L1448 IRS2, performed with $N$=100 and $n$=1000. Same as Figure \ref{fig:obs_b335_KS}.
}
\label{fig:obs_L1448_KS}
\end{figure*}

\begin{figure*}[!tbh]
\centering
\includegraphics[scale=0.5,clip,trim= 3cm 2cm 2cm 0cm]{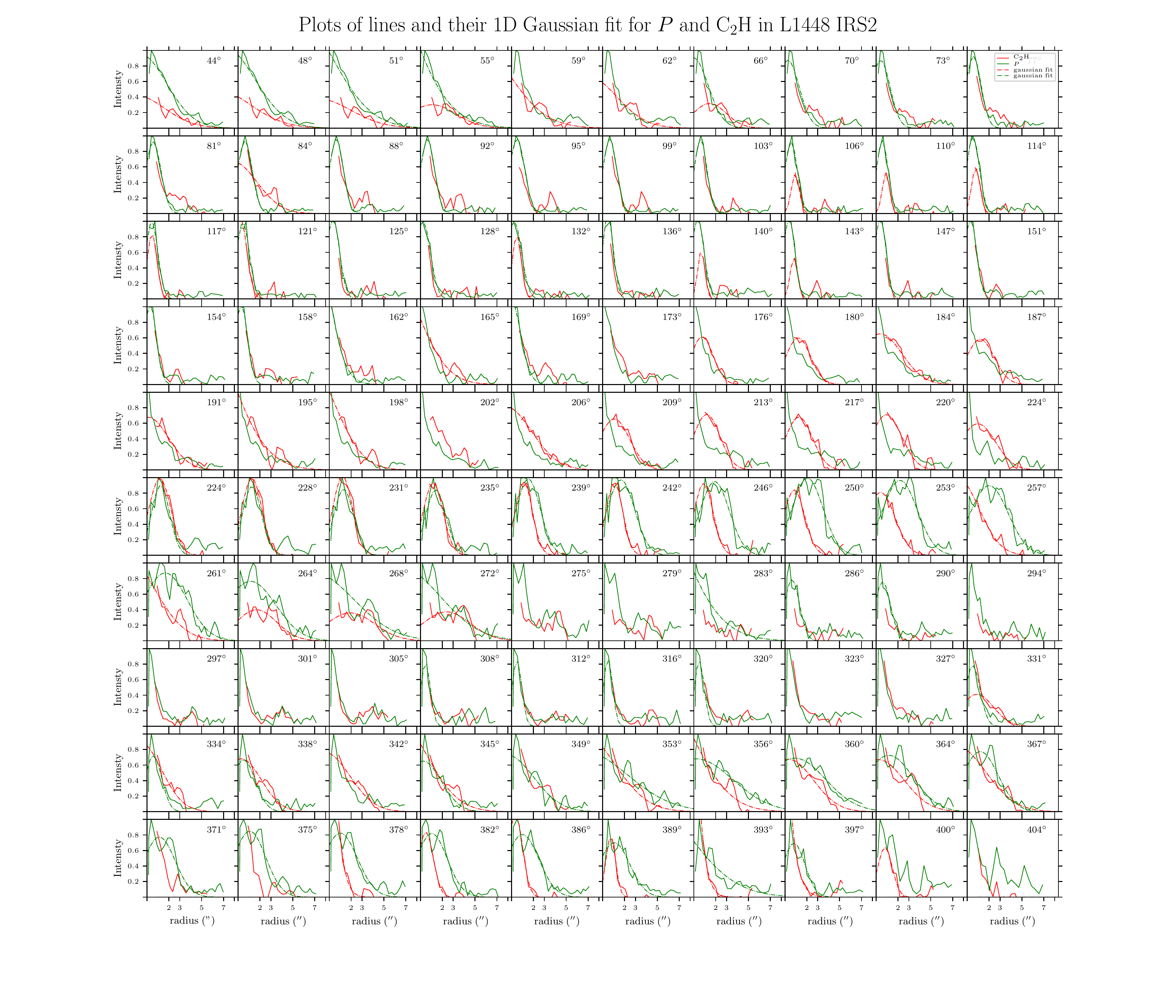}\caption{\small Emission profiles of the S/N selected lines obtained in the C$_2$H moment 0 and $P$ maps of L1448 IRS2. Same as Figure \ref{fig:obs_b335_1D}.}
\label{fig:obs_L1448_1D}
\end{figure*}

Figures \ref{fig:obs_L1448_KS} and \ref{fig:obs_L1448_1D} present the same tests presented in Section \ref{sec:p3_obs_KS}, but applied to L1448 IRS2, using $N$=100 and $n$=1000. This time the departure of the lines is offset by $\sim$ 50 au from the peak of the total intensity map. We chose this because the central region is thought to be optically thick and the corresponding polarized emission to be contaminated by self-scattering \citep{Kwon2019}. As in B335, the KS test results suggest that the two distributions ($\sigma$,$\mu$)  between the C$_2$H moment 0 and $P$ maps in L1448 IRS2 could come from a common parent distribution. Therefore, in this core  the morphology of the emission in these two maps could also have a common origin.

\noindent\begin{minipage}{\textwidth}\centering
\includegraphics[scale=0.5,clip,trim= 3cm 2cm 3cm 0cm]{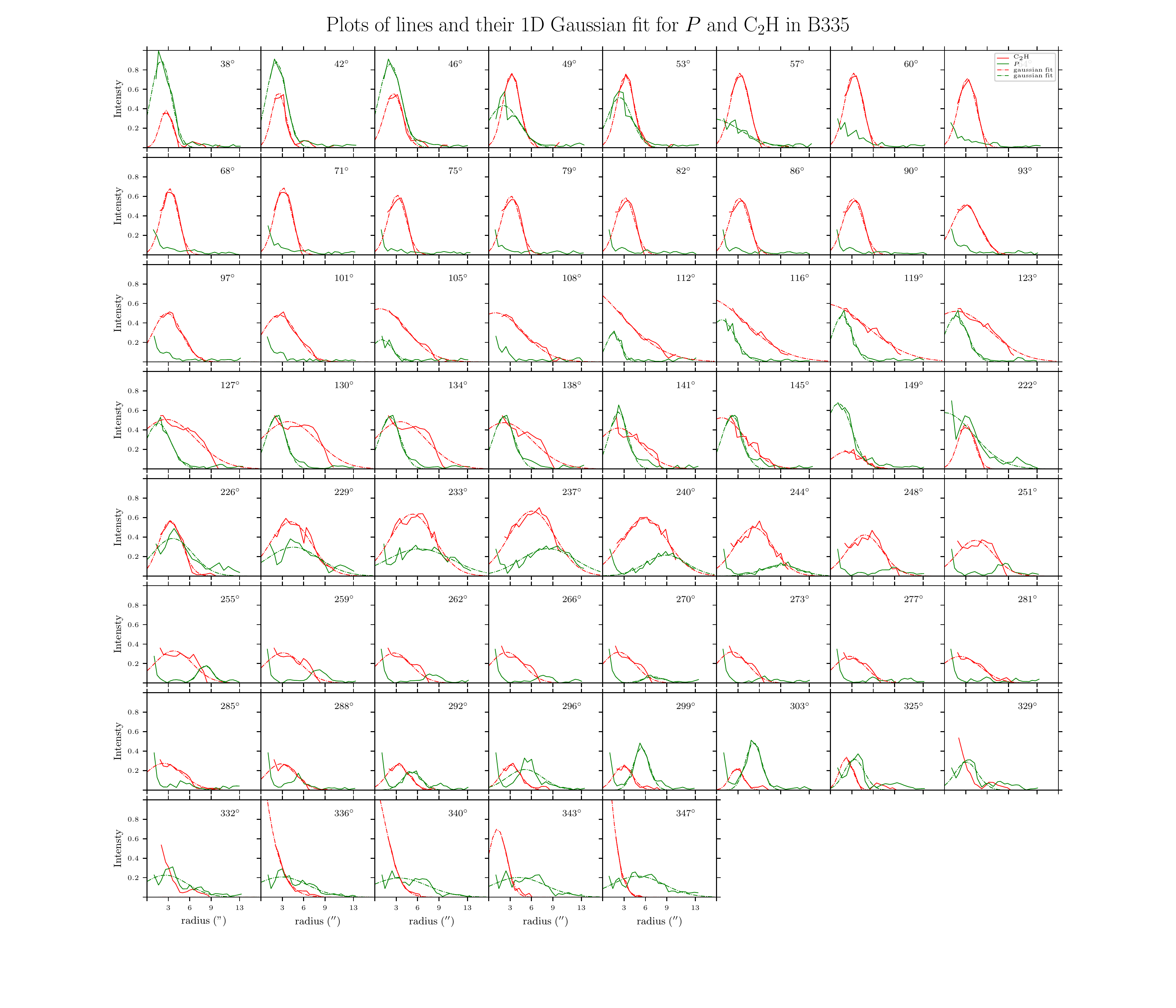}
\captionof{figure}{\small Emission profiles of the S/N selected lines obtained in the C$_2$H moment 0 and $P$ maps of B335. The red and green solid line corresponds  to the emission profile along a given line of the C$_2$H moment 0 and $P$ map, respectively, shown in Figure \ref{fig:obs_b335_KS}. The angle in each panel is the position angle (taken from the north) of the line (see Figure \ref{fig:obs_b335_KS}). When the 1D Gaussian fit converges (for the white lines in Figure \ref{fig:obs_b335_KS}), it is shown as a dashed line.
}
\label{fig:obs_b335_1D}
\end{minipage}

\end{appendix}
\end{document}